\documentclass{article} 

\usepackage[square]{natbib}
\renewcommand{\cite}{\citep} 
\usepackage{citesort} 

\title{ Empirical Equivalence, Artificial Gauge Freedom and a Generalized Kretschmann Objection }  
\author{J. Brian Pitts \\ University of Notre Dame}

\begin{document}
       
\maketitle

 \begin{abstract}  

	Einstein considered general covariance to characterize the novelty of his General Theory of Relativity (GTR), but Kretschmann thought it merely a formal feature that any theory could have.   The claim that GTR is ``already parametrized'' suggests analyzing substantive general covariance as formal general covariance achieved without hiding preferred coordinates as scalar ``clock fields,'' much as Einstein construed general covariance as the lack of preferred coordinates. Physicists often install gauge symmetries artificially with additional fields, as in the transition from Proca's to Stueckelberg's electromagnetism.  Some post-positivist philosophers, due to realist sympathies, are committed to judging Stueckelberg's electromagnetism distinct from and inferior to Proca's.  By contrast, physicists identify them, the differences being  gauge-dependent and hence unreal.  It is often useful to install gauge freedom in theories with broken gauge symmetries (second-class constraints) using a modified  Batalin-Fradkin-Tyutin (BFT) procedure.  Massive GTR, for which parametrization and a Lagrangian BFT-like procedure appear to  coincide, mimics GTR's general covariance apart from telltale clock fields.  A generalized procedure for installing artificial gauge freedom subsumes parametrization and BFT, while being more Lagrangian-friendly than BFT, leaving any primary constraints unchanged and using a non-BFT boundary condition.  Artificial gauge freedom licenses a generalized Kretschmann objection.  However, features of paradigm cases of artificial gauge freedom might help to demonstrate a principled distinction between substantive and merely formal gauge symmetry.

 \end{abstract} 


\section{Introduction}

 It has been claimed, starting with Einstein in the 1910s, that general covariance is the chief novel and distinctive feature of the General Theory of Relativity (GTR), the great lesson about space-time physics that all theorizing in the foreseeable future ought to respect.  Whatever the details might be, general covariance is supposed to involve the absence of certain structures that are  impervious to the contents and history of the world. Initially Einstein took general covariance to be manifested in the admissibility of arbitrary coordinate systems in GTR.  However,  almost immediately Erich Kretschmann cast doubt on Einstein's analysis of general covariance and argued that it was a merely formal feature of GTR that  any theory (of the sort that physicists might reasonably entertain), including GTR's predecessors and competitors, could share if formulated with a bit of ingenuity  \cite{Kretschmann,NortonStumble,RynoKretschmann,NortonGC}.  
After conceding the truth of Kretschmann's criticism that general covariance could be  a merely formal property reflective of  the theorist's 
ingenuity, Einstein then maintained that GTR was (substantively) generally covariant in the special sense that the field equations of GTR took their simplest form when expressed (formally) generally covariantly, thereby implying that other theories would look more complicated  in that form. 
Thus Einstein in effect distinguished formal or weak general covariance from substantive or strong general covariance, if one may use terms that would arise later \cite{BergmannLectures,StachelGC}.  
While something seems intuitively right about Einstein's claim, the task of analyzing or replacing the ``simplest form'' criterion for substantive general covariance thus far has not received a universally satisfying resolution.  
 Although it is widely believed that there is some  unique nontrivial notion of general covariance that at least roughly captures the main innovation of Einstein's 1915-6 theory of gravity, perhaps matters are not so simple. Elsewhere I have discussed the Anderson-Friedman absolute objects program and also a variational criterion as candidate analyses \cite{FriedmanJones,PittsPhilDiss}. The Anderson-Friedman project appears to fail if one expects that GTR should count as substantively generally covariant, on account of certain kinds of geometric objects that are susceptible to absoluteness simply by being nonzero everywhere \cite{FriedmanJones,PittsPhilDiss,GiuliniAbsolute,ZajtzGerm}.  In particular the volume element $\sqrt{-g}$ counts as absolute.  A variational analysis might be extensionally correct, but there is some concern that it might give the right answer for the wrong reason \cite{PittsPhilDiss}.

A  suggestion sometimes made by physicists about GTR, that it is an ``already parametrized theory,'' provides another fruitful way to analyze general covariance. Parametrization involves the use of preferred coordinates as scalar fields, known as ``clock fields''  
  \cite{Lanczos49,BergmannBrunings,ADMparameter1,ADMcanonical,ADMdynamics,DiracLQM,Kuchar73,TeitelboimCommutators,Kuchar81,TorrePara,Westman,LopezGotayMarsden}.
A parametrized theory will have, in addition to the fields that one might have expected, a set of $n$ scalar fields $X^A$ (in $n$ space-time dimensions) which appear (often differentiated) in the action principle and equations of motion.  For example, a parametrized special relativistic theory will have clock fields in the Poincar\'{e}-invariant combination
$\eta_{\mu\nu} =_{def} X^A,_{\mu} \eta_{AB} X^B,_{\nu},$ where $\eta_{AB}=diag(-1,1,1,1)$ and the comma denotes differentiation with respect to arbitrary coordinates $x^{\mu}$; the flat metric tensor is reduced to a function of (derivatives of) the clock fields.    Clock fields have recently come to the attention of some philosophers of physics, such as John Norton, John Earman and Gordon Belot  \cite{NortonGC,EarmanUGR,EarmanCovariance,BelotPoP}.  The idea behind the phrase ``already parametrized'' is that GTR does not need any explicit clock fields because they are already there, albeit obscured and mixed in with the true degrees of freedom.  Hopes for digging out the clock fields from the metric by a canonical transformation, or even regarding GTR as literally parametrized without exhibiting the clock fields explicitly, have diminished somewhat over the years.  However, whether clock fields are already in GTR somewhere does not matter for my purpose of analyzing general covariance.  What matters is that no further clock fields are needed for GTR to achieve formal general covariance.   
At present it appears that the absence or presence of (non-redundant \cite{BergmannBrunings}) clock fields  in a formally generally covariant theory  correctly tracks expectations that a theory formulation is or is not substantively generally covariant, respectively,  assuming that all non-variational fields (that is, fields present but not varied in the action \cite{GotayIsenberg}) have been replaced by functions of the clock fields and their derivatives.   Clock fields are  preferred coordinates in disguise, so the criterion that a formally generally covariant theory formulation is substantively generally covariant just in case it lacks (non-redundant) clock fields  revives the spirit of  Einstein's mid-1910s claim that general covariance involves the lack of preferred coordinate systems. Most people nowadays agree that there is nothing very interesting about coordinate systems \cite{EarmanCovariance}, so it is surprising that so old-fashioned a criterion turns out to work so well, and indeed perhaps better than the geometrically  sophisticated Anderson-Friedman project.   The absence of clock fields and the absence of fields not varied in the action principle seem to be not merely coextensive, but very simply related conceptually.   The functional form  of a parametrized theory's dependence on the clock fields determines the theory's symmetry group in a simple way.

For typical gauge theories such as Maxwell's electromagnetism, the Yang-Mills theories of the weak and strong nuclear forces, and Einstein's theory of gravity, the presence of gauge freedom follows largely from the locality of the field theories, the Poincar\'{e} invariance of Special Relativity, and the presence of a pure spin particle(s) (spin $1$, several spin $1$, and spin $2$, respectively) with no lower-spin admixture.  Lower spin admixtures in special relativistic field theories tend to give wrong-sign degrees of freedom, such as the time component of a $4$-vector, the time-space components of a symmetric rank $2$ tensor, and the like \cite{Wentzel}.  Such negative-energy degrees of freedom can threaten stability, especially in the presence of interactions.  For electromagnetism, the massive Proca theories avoid gauge freedom.  One might expect similar results for massive variants Yang-Mills theories and GTR, but subtle difficulties arise \cite{DeserMass,UnderdeterminationPhoton}.

 The question whether there exist empirically equivalent but incompatible  theories bears on how tightly empirical constraints from the progress of science might constrain our theorizing.  While the issue of empirical equivalence has been widely discussed, philosophers' discussions often have involved rather thin examples, perhaps generating new theory candidates by de-Ockhamizing (replacing one theoretical entity, perhaps ``force,'' by some  combination of multiple entities such that only that combination plays a role in the theory, such as ``gorce plus morce'' \cite{GlymourEpist}), 
\emph{ad hoc} deletion of some  regions of space-time or objects therein while the remainder behaves just as in the mother theory,  
and the like.  It will turn out, surprisingly, that some physically interesting resembles bear some resemblance to de-Ockhamization and that physicists have elaborate technologies for performing such reformulations.  
Using physically interesting examples  also has the advantage that fairly detailed principles for theory identification are already suggested by standard  techniques and attitudes in physics. Thus there is comparatively little risk that the question of theoretical equivalence will be resolved one way or another simply by  \emph{ad hoc} stipulation. 
Using examples from real physics  avoids collapsing scientific underdetermination into familiar extrascientific skeptical worries, as P. Kyle Stanford has urged \cite{StanfordUnconceived}.  John Earman recently has made some surprising claims that certain empirically equivalent theory formulations amount to different theories \cite{EarmanCovariance}; these claims will be addressed below.

Empirical equivalence is an issue that plays a key role in arguments about scientific realism.  According to Andr\'{e} Kukla, 
\begin{quote}
[t]he main argument for antirealism is undoubtedly the argument from the underdetermination of theory by all possible data.  Here is one way to represent it: (1)  all theories have indefinitely many empirically equivalent rivals; (2) empirically equivalent hypotheses are equally believable; (3) therefore, belief in any theory must be arbitrary and unfounded. \cite[p. 58]{Kukla}. \end{quote} %
During the heyday of logical empiricism, many influential people   denied that distinct and incompatible  but empirically equivalent theories existed \cite{GlymourTheoretical}. In particular,  Carnap and Reichenbach had no room empirically equivalent theories, given the verificationist criterion of meaning \cite{ReichenbachEP,CarnapMetaphysics,Muhlholzer}. 
 But not only W. V. O. Quine's work \cite{QuineEquivalent},
 but also the revival of scientific realism during the 1960s-70s \cite{GroverMaxwell}, led to a revival of belief that distinct and incompatible but empirically equivalent theories exist.  
More recently the view that there do exist rival empirically equivalent theories has been somewhat  widely held \cite{MusgraveRealism,KuklaEquivalence,EarmanCovariance}, in contrast to the earlier positivist view that empirically equivalent theories say the same thing and so are merely linguistic variants. 
This work and a companion piece \cite{UnderdeterminationPhoton} aspire to address the question of empirically equivalent theories within the context of local classical and (to some degree) quantum field theories.  

The longstanding issues of  general covariance and empirical equivalence  are strongly tied to the largely untouched conceptual and technical question of artificial gauge freedom. 
Gauge freedom here is  broadly construed as a theory formulation's having arbitrary functions of space-time in the solution of the equations of motion. Given a suitable theory formulation that violates general covariance on some reading or that lacks a gauge symmetry, it is often possible to construct an empirically equivalent formulation that satisfies that analysis or has a gauge symmetry.  The most famous and systematic procedure for installing artificial gauge freedom  is the Batalin-Fradkin-Tyutin procedure (BFT) \cite{BF86,BF87,BFF89,BT91}, 
which takes constrained Hamiltonian theory formulations  with broken gauge symmetries  (expressed technically as ``second class constraints'') as input and yields formulations with unbroken gauge symmetries (expressed technically as ``first class constraints'') as output.   The philosophically interesting spirit can be generalized, as will appear below, not only to Lagrangian formulations of theories, but to theories that have \emph{no} constraints.  Given a field theory, one can add additional fields and additional symmetries (some examples of which will be more interesting than others) to get an empirically equivalent theory formulation.  Thus the BFT procedure and similar techniques provide an algorithm for either constructing formulations exemplifying the Kretschmann-type worry that general covariance or gauge freedom is a merely formal rather than substantive feature of a theory formulation, or for generating empirically equivalent theories.

Are the results physically interesting examples of empirically equivalent but incompatible theories? Or are the results merely alternative formulations of the same theory, as physicists would expect?  Making the most natural application to electromagnetism, the question becomes:  is Proca's non-gauge massive electromagnetism (with some definite photon mass) really the same theory as Stueckelberg's gauge massive electromagnetism (with the same photon mass), or are they different theories? The two theory formulations are empirically equivalent, but might seem \emph{ontologically} distinct. Proca formulations have four fields and three degrees of freedom at every spatial point, with no gauge freedom, whereas Stueckelberg formulations have five fields and three degrees of freedom at each spatial point, with gauge freedom.  Clearly $4 \neq 5$ and the absence of gauge freedom differs from the presence of gauge freedom, so there is at least some temptation to regard the theories as distinct.   The presence of gauge freedom is often said by particle physicists to be of fundamental significance.  Even if these claims are overstated (as Chris Martin argues with considerable plausibility \cite{CMartinGaugeArgument}), it is noteworthy  that gauge freedom   by itself fails to distinguish Maxwell's massless theory from the Stueckelberg formulation of massive electromagnetism.  Empiricist-minded philosophers might share with most physicists the view that Proca's and Stueckelberg's families of theory formulations are really the same family of theories and that a Proca electromagnetism with a given mass and a Stueckelberg electromagnetism with that same mass are the same theory.   The fact that one can gauge-fix Stueckelberg formulations to produce Proca formulations provides a technical implementation of the empirical equivalence of the two families.  Alleged ontological differences that cannot have any empirical consequences and that disappear on gauge-fixing are not real, one might think.  But then the presence or absence  of gauge freedom in massive electromagnetism is not a real feature of the theory, but merely a conventional choice of formulation. 
Thus massive electromagnetism is subject to a Kretschmann-style point:  gauge freedom, if not presently initially, can be installed with a little effort.   In installing artificial gauge freedom, I have in mind especially the classical theory formulated in Lagrangian terms.  Quantization can make an important difference, even rendering a classically satisfactory theory unacceptable under quantization, due to failure of unitarity or renormalizability, as is generally believed to be the case of most massive Yang-Mills theories.  Essentially Abelian theories with a mass term for the photon only \cite{UnderdeterminationPhoton} are probably an exceptional case for which the mass term causes no difficulty.  No such trouble arises for electromagnetism, however.



Whereas the Stueckelberg trick (as it is often called) for installing gauge freedom into massive electromagnetism is basically an \emph{ad hoc} or opportunistic move, it has become clear in recent decades that there is a general systematic procedure that achieves basically the same result.  This is the  Batalin-Fradkin-Tyutin \emph{et al.} framework \cite{BF86,BF87,BFF89,BT91}. 
 The BFT procedure in fact encompasses various tricks  that are employed in standard examples.  For massive Yang-Mills fields (not just the degenerate electromagnetic special case), 
 Stueckelberg's trick \cite{Ruegg} takes a massive theory with broken gauge symmetry due to the mass term, and then restores the symmetry with the introduction of one or more compensating fields.  

The BFT procedure, strictly construed, has some features that make application to massive gravity an unpleasant prospect, but some simple modifications suggested below yield a procedure that apparently agrees with parametrized massive gravity with four clock fields.
Below I call attention to some disadvantages and limitations of the existing BFT technologies for installing artificial gauge freedom, and outline simpler procedures of both Lagrangian and Hamiltonian varieties.  The procedures are illustrated using Proca's massive electromagnetism, where they recover the usual Stueckelberg gauged massive electromagnetism rather more directly than does BFT. It is inconvenient that, due to the `boundary condition' imposed on the new gauge compensation fields, the usual BFT procedure requires a canonical transformation to restore the typical velocity-momentum relationship and thus recover the Stueckelberg Lagrangian density for massive gravity.  An alternative Lagrangian-friendly boundary condition is proposed below.  

 I also outline a more general strategy of installing artificial gauge freedom. The BFT procedure adds extra fields only to convert theories with second-class constraints\footnote{Constraints  vanish for all dynamically possible trajectories and arise due to degenerate kinetic terms that obstruct the Legendre transformation from velocities to momenta.  The term, somewhat confusingly, refers both to the expression that vanishes for dynamically possible trajectories (but not for all kinematically possible trajectories) and to the vanishing of that expression.  Taking constraints as expressions of the phase space variables (and momentarily forgetting that they have the value of zero), one can take Poisson brackets of pairs of constraints (and then set the values to zero).  Second class constraints have nonzero Poisson brackets with at least some   other constraint(s). First class constraints have Poisson brackets that vanish, perhaps with terms proportional to the constraints themselves,  with all the constraints \cite{Sundermeyer}.}  into gauge theories with only first class constraints. Furthermore, the BFT procedure converts all the second class constraints into first class constraints.  But one might wish to convert only some second class constraints into first class constraints.  More importantly, one might wish to add gauge freedom to theories with no constraints, such as by parametrizing a scalar field theory \cite{Kuchar73,Kuchar81,RovelliBook}. Perhaps one could have both of these goals. In at least some cases, such goals can be realized with the addition of extra fields and gauge symmetries.

  Knowing of BFT-type procedures, one can show that certain examples of general covariance or gauge freedom are artificial and formal; this result  might help to determine which, if any, examples are natural and substantive.  Standard gauge-fixing techniques in effect reverse the BFT procedure and so suggest the theoretical equivalence of pre-BFT and post-BFT formulations.  Induction over telltale signs of BFT-type installation of artificial freedom might help to answer the generalized Kretschmann objection.


\section{Parametrization: Preferred Coordinates as Clock Fields and the Einstein-Kretschmann Debate}

   As Earman notes, clock fields threaten the criterion of no nonvariational fields as the test for general covariance \cite{EarmanCovariance}.  While Earman declines to seek a more refined criterion to exclude theories with clock fields, it seems to me that the relevant lesson here is that nonvariational fields can be converted into functions of clock fields and their derivatives.  If nonvariational fields can be reduced to clock fields, then checking for nonvariational fields while permitting clock fields is a criterion bound to fail:  one can simply convert any erstwhile nonvariational fields into functions of clock fields and their derivatives  to get an equivalent formulation lacking nonvariational fields, thus satisfying the condition for substantive general covariance even in clearly absurd cases such parameterized field theories in flat space-time.  Such a test for general covariance would be like a police search for criminals in a house without guarding the back door.  

It therefore seems that a promising analysis of the assumed substantive general covariance of GTR can be found in the claim  that GTR is ``already parameterized''   \cite{ADMparameter1,ADMparameter2,ADMcanonical,ADMdynamics,DiracLQM,KucharBubble,Kuchar73,Kuchar81,KucharHyperI,KucharHyperII,KucharHyperIII,Kuchar78,TorreNotPara}.
Whereas field theories in Minkowski space-time (or other pre- or non-generally covariant theories) admit parameterized formulations with the explicit introduction of clock fields to serve as preferred coordinates in disguise,
and nonrelativistic mechanics admits treating time as a dynamical variable in terms of some new parameter \cite{Lanczos49}, such theories do not naturally come in this parameterized form.  Originally there was hope that GTR was quite literally already parameterized, with clock fields built into the theory  in such a fashion that, with sufficient cleverness (likely involving a nonlocal canonical transformation, use of coordinates based on scalar concomitants of the Riemann tensor, \emph{etc.}) one might explicitly identify the clock fields and deparametrize the theory.  Parameterized theories display important formal similarities to GTR in terms of certain Poisson bracket relationships that tend to strengthen the impression that GTR is already parameterized \cite{DiracLQM,IshamKucharI}.  It turns out that
there are certain technical obstacles that exclude the strongest versions of the claim that GTR is already parametrized \cite{TorreNotPara,Westman}, but there might be plausible senses in which the claim is true.  Given the analogy between parametrized field theories and GTR, there has been considerable work on parametrized field theories as a step towards quantum gravity \cite{ADMparameter1,ADMparameter2,ADMcanonical,ADMdynamics,TorreNotPara,DiracLQM,KucharBubble,Kuchar73,Kuchar81,Kuchar78,TorrePara,LeeWald,HajicekKuchar,KucharStoneParam,IshamKucharI,Sundermeyer,ChoVaradarajan,Varadarajan,Varadarajan2,BelotPoP}.  Some early works in quantum gravity employed clock fields even in GTR \cite{BergmannBrunings,BergmannPSZ}, but this approach did not persist.  Given the research trends, one might conclude that, whether GTR is already parametrized in some deep sense or not, at least GTR is distinctive in admitting arbitrary coordinates without the need to \emph{install} clock fields.

In the typical case of a field theory in Minkowski space-time, the parametrization process (in its simpler Lagrangian version) proceeds as follows.  One begins with the Lorentz vectors, tensors, \emph{etc.} (here written with capital Latin letters $A,$ $B$ \ldots for the indices, such as $\eta_{AB}$), covariant under the Poincar\'{e} group.  One then transmogrifies them into world (ordinary, coordinate)  vectors, tensors, \emph{etc.} (here written with small Greek indices $\mu,$ $\nu$, \ldots), thus defining their components for arbitrary coordinate systems using the relevant tensor transformation law.  For the flat background metric, one now has not the matrix $diag(-1,1,1,1)$ but a  Riemann-flat tensor 
\begin{equation}
\eta_{\mu\nu} = X^{A},_{\mu} \eta_{AB} X^{B},_{\nu}, \end{equation} where $X^A$ is a set of Cartesian coordinates,  $x^{\mu}$ is a set of arbitrary coordinates, and the comma indicates partial differentiation.  
 Thus the Lorentz metric $\eta_{AB}$ is replaced by a flat metric tensor $\eta_{\mu\nu},$  partial derivatives with respect to Cartesian coordinates are replaced with $\eta$-covariant derivatives, and so on. For typical dynamical fields,  little change is needed beyond replacing capital Latin letters with small Greek letters, the installation of $\eta$-covariant derivatives, and a (perhaps trivial)  choice of density weight (that distinction being almost trivial at the Lorentz covariant level  because Lorentz transformations have (anti)unit determinant). If an electromagnetic (co)vector potential potential $A_{B}$ is present, the prudent choice (though not the only possible one) is to change it into a coordinate covector $A_{\mu}$ (with zero density weight).  The Lagrangian density, a Lorentz scalar, must become a weight $1$ scalar density to make the action a coordinate scalar \cite{Anderson}. For spinors, which might never have been  discussed in this context, this process  is nontrivial; in principle the OP-Bilyalov formalism should suffice \cite{OPspinor,IshamSalamStrathdee,ChoFreund,BilyalovSpinors}, but with strong nonlinearity in the clock field gradients and some mild inequalities restricting the coordinates. 
The key step in parametrization, after arbitrary coordinates have been introduced as above, is to treat the nonvariational fields such as $\eta_{\mu\nu}$ not as primitive, but as derived according to the definition 
\begin{equation}
\eta_{\mu\nu} =_{def} X^{A},_{\mu} \eta_{AB} X^{B},_{\nu}. \end{equation} 
Now the dynamical fields such as the electromagnetic vector potential $A_{\mu}$ are tensorial, and the clock fields $X^A$ with a typically nontrivial relation $X^{A}(x^{\mu})$ to the arbitrary coordinates  $x^{\mu}$ achieve formal general covariance.
The presence of the matrix $\eta_{AB}$ contracted with the clock field gradients serves as a reminder that the theory is Poincar\'{e}-covariant in terms of the preferred coordinates $X^A$.   (For metrics with nonzero constant curvature, there will be dependence on one or more of the clock fields themselves, not just the derivatives.) 
Then one promotes, or perhaps rather demotes \cite{Kuchar73}, the preferred coordinates $X^A$ into variational fields in the action principle, which is expressed using the arbitrary coordinates  $x^{\mu}$. Now the clock fields and the dynamical fields enter the action $S[A_{\mu}, X^B]$ in basically the same fashion, with both $A_{\mu}$ and $X^B$ having Euler-Lagrange equations 
\begin{eqnarray} 
\frac{\delta S}{\delta A_{\mu} }=0,  \frac{\delta S}{\delta X^B}=0 \end{eqnarray}
and being functions of the arbitrary coordinates $x^{\mu}.$  (The use of electromagnetism is merely illustrative.)   While much of the work on parametrized field theories occurs in a Hamiltonian formalism, the Lagrangian formalism is appropriate for my purposes and leaner in field variables by doing without momenta. 

 One readily demonstrates the claim  that it does not much matter whether $X^A$ are varied or not, because the Euler-Lagrange equations from varying $X^A$ are identical to the result of using the other fields' Euler-Lagrange equations in the generalized Bianchi identities.\footnote{Generalized Bianchi identities are (typically differential) identities that hold among the Euler-Lagrange derivatives of the fields in the action principle in cases where the action is invariant are transformations described by arbitrary functions \cite{Sundermeyer}.} 
  Let the other fields be represented by a weight $w$ $(1,1)$ tensor density  $\phi^{\alpha}_{\beta}$ \cite{Schouten,Anderson,Israel}. This case is representative of all tensor fields.  (Nothing crucial would change if a connection or spinor field were included, though the details would vary.)  Because the action $S$ is a scalar, it is unchanged under arbitrary infinitesimal changes of the coordinates $x^{\mu},$ including the ones generated by an arbitrary vector field $\xi^{\mu}.$   Taking this coordinate transformation to have compact support makes all boundary terms to disappear, including those from pulling off derivatives from  $\xi^{\mu}$ in $\pounds_{\xi} \phi^{\alpha}_{\beta} = \xi^{\mu} \phi ^{\alpha}_{\beta},_{\mu} - \phi^{\mu}_{\beta} \xi^{\alpha},_{\mu} + \phi^{\alpha}_{\mu} \xi^{\mu},_{\beta} + w \phi^{\alpha}_{\beta} \xi^{\mu},_{\mu}. $ After making some rearrangements and using the arbitrariness of $\xi^{\mu}$ to pull off the integral sign, one obtains the generalized Bianchi identity
\begin{equation}
\phi^{\alpha}_{\beta}, _{\mu} \frac{\delta S}{\delta \phi^{\alpha}_{\beta} } - w \frac{\partial}{\partial x^{\mu} } \left(  \phi^{\alpha}_{\beta}  \frac{\delta S}{\delta \phi^{\alpha}_{\beta} }  \right)
+ \frac{\partial}{\partial x^{\alpha} } \left(   \phi^{\alpha}_{\beta}  \frac{\delta S}{\delta \phi^{\mu}_{\beta} }  -  \phi^{\beta}_{\mu}  \frac{\delta S}{\delta \phi^{\beta}_{\alpha} }\right) + X^A,_{\mu} \frac{\delta S}{\delta X^A } = 0.
\end{equation}
Letting $\phi^{\alpha}_{\beta}$ be on-shell (that is, letting its Euler-Lagrange equations hold) leaves 
\begin{equation}
X^A,_{\mu} \frac{\delta S}{\delta X^A } = 0.
\end{equation}
Clock fields have non-vanishing linearly independent gradients, so \begin{equation} \frac{\delta S}{\delta X^A } =0, \end{equation}  which was to be demonstrated.  
Nothing important changes if second (or higher) derivatives of $X^A$ are present, as they will be if a background connection is reduced to a function of clock fields and their derivatives.

Here the explicit form of the fields besides $X^A$ ultimately does not matter, but the simple form of the Lie derivative of scalar fields 
\begin{equation}
\pounds_{\xi} X^A = \xi^{\mu} X^A,_{\mu},
\end{equation} which is algebraic in the vector field  $\xi^{\mu}$ generating the coordinate transformation, does the important work.  The fact that the background structure (as picked out by nonvariational fields in the original action) is reduced to a function of \emph{scalar} fields and their derivatives immunizes  clock fields against the major difficulty for the Anderson-Friedman program.  That program seems to be thwarted by the fact that certain kinds of geometric objects, including scalar densities of any nonzero weight and contravariant (tangent) vectors and contravariant vector densities of any weight except $1$, satisfy the absoluteness condition of local diffeomorphic equivalence just by virtue of being nonzero \cite{Anderson,FriedmanFoundations,ZajtzGerm,FriedmanJones,PittsPhilDiss}.\footnote{Some relevant results in differential geometry in (\cite{PittsPhilDiss}) were first obtained and supplied by Robert Geroch in relation to the Anderson-Friedman program (private communication); later it was found that Zajtz had anticipated the results but without reference to the Anderson-Friedman program \cite{ZajtzGerm}.}  Thus the fact that one can set $\sqrt{-g}$ to $1$ in a neighborhood for any metric in any space-time \cite{FriedmanJones,GiuliniAbsolute}, for example, makes this an apparently false positive test for absolute objects, in that GTR itself counts as having an absolute object and hence not substantively generally covariant.   Likewise the fact that one can set any (nowhere vanishing!)   tangent vector field to have components $(1,0,0,0)$ in a neighborhood of any point means that one cannot use such a test to distinguish some such fields as absolute and others as dynamical.  (The nowhere vanishing condition implies that the Jones-Geroch dust counterexample fails, because dust can have holes.  A different strategy is required for spinors \cite{FriedmanJones}, for which it is typical, but unnecessary \cite{OPspinor,BilyalovSpinors}, to introduce an orthonormal basis.) 
The coordinate transformation laws (of which the Lie derivative gives the infinitesimal form) for different kinds of geometric objects are various and in some cases have the surprising consequence that all such fields are locally alike;
I have called such geometric objects ``susceptible to absoluteness.''  For  geometric objects susceptible to absoluteness, such as (nowhere vanishing) tangent vectors or scalar densities,  one cannot using local sameness up to coordinate transformations to distinguish some as absolute and some as dynamical.  This difficulty for the Anderson-Friedman program \cite{FriedmanJones,GiuliniAbsolute} might be fatal.  The use of clock fields as the test for lack of substantive general covariance, however, relies only on the properties of scalar fields, the simplest type of geometric object.  It is difficult to imagine that scalar fields can hold similar nasty surprises; they do not even have a nontrivial transformation law.    In the absence of any analogous reason to expect  false positives from clock fields, clock fields seem like a good criterion for the lack of substantive general covariance.

The functional form  of a parametrized theory's dependence on the clock fields determines the theory's symmetry group in a simple way, as a typical example illustrates sufficiently.     The  translation subgroup  of the Poincar\'{e} group manifests itself in having only differentiated clock fields in the formulation, because $(X^{A} + c^A),_{\mu} = X^{A},_{\mu}$ for constants $c^{A}$.  The Lorentz subgroup $O(1,3)$ manifests itself in having only the Lorentz-invariant combination of the clock field gradients $X^A,_{\mu} \eta_{AB} X^B,_{\nu}$ in the theory formulation, not each gradient by itself.  Not coincidentally, the Poincar\'{e} group is just the group of Killing vector fields for a flat metric tensor $\eta_{\mu\nu}.$

Of late philosophers of physics have taken to discussing parametrized theories \cite{NortonGC,EarmanUGR,EarmanCovariance,BelotPoP}.  As John Norton discusses, Einstein often ascribed to inertial coordinate systems the (objectionable to Einstein) property of acting but not being acted upon; even late in life, when the concept of geometric object would have been available \cite{SchoutenHaantjes,Nijenhuis}, Einstein persisted in faulting inertial coordinate systems  \cite{NortonTriumph}.  
It might be tempting to regard Einstein's views as outmoded, and presumably surpassed by more sophisticated projects such as that of Anderson and Friedman.    As John Earman notes, it is widely held that there is nothing very interesting about coordinate systems \cite{EarmanCovariance}.  Earman then proceeds to discuss clock fields as an obstacle to defining substantive general covariance, using the example of unimodular gravity.  He  declines to suggest yet another notion of general covariance  to avoid this  problem.

My proposal that the absence of clock fields be used as the test for substantive general covariance can be viewed as taking up Earman's challenge.  Whereas he ultimately decides that appeals to the notion of gauge-invariant observables is promising and that further search for a formal criterion for substantive general covariance is unpromising, both claims are questionable.  Concerning the   use of the Rosen-Sorkin Lagrange multiplier trick  \cite{RosenMultiplier,Rosen73,SorkinScalar,EarmanCovariance,GiuliniAbsolute}  to produce a formally generally covariant scalar field theory in Minkowski space-time,  Earman judges the resulting formulation to be a distinct theory from the more usual scalar field theory, to which the former reduces on gauge-fixing and with which it is empirically equivalent.  This surprising claim that two empirically equivalent scalar field theory formulations are inequivalent theoretically is based on their having different sets of observables, in a technical physical sense \cite{EarmanCovariance} (presumably in the  sense discussed long ago by Peter Bergmann \cite{BergmannLectures,Bergmann}, though the notion of ``observable'' in the literature is not obviously unified \cite{Sundermeyer,HealeyMcTaggart,Lusanna2,PonsSalisburySundermeyer}). 
 Many physicists would have strong intuitions that these pairs of theories are each equivalent (though that judgment is not universal \cite{GiuliniAbsolute}), largely because Lagrange multipliers are seen as just a trick to allow the variation of something that doesn't fundamentally need to be varied in an action principle (especially if the Lagrange multiplier is a geometric object by itself, as opposed to the time component of a covector as in Maxwell's electromagnetism, for example).  As Earman himself says \cite[p. 3]{EarmanCovariance}, widely shared intuitions among physicists should count as data for philosophers of physics, even if such intuitions are not decisive.  If distinct empirically equivalent theories are so easily had as Earman suggests, one might worry about the status of some quantum field theories, such as massive quantum electrodynamics, in which, intuitively, a single theory is proven to be unitary with one gauge-fixing and renormalizable with a different gauge-fixing \cite[pp. 738, 739]{PeskinSchroeder}.  It might be worrisome if intuitively a single quantum field theory proved on reflection to bifurcate into two distinct theories, one of which is provably renormalizable but perhaps not unitary, another of which is unitary but perhaps not renormalizable. 

  The question arises whether one should take having different technical observables for empirically equivalent theories to shed light on theoretical equivalence, or use it to criticize the  notion of observables that Earman employs.   
Peter Bergmann, one of the key authors concerning this concept of observables, regarded as trivial the weak general covariance on which Earman's point largely rests \cite{BergmannLectures}. 
The point of Bergmann's demanding gauge invariance for observables is to ensure \emph{determinism} for their time evolution in the context of GTR; otherwise \emph{predictability} is lost because changing from one coordinate description to another \emph{equally valid} description after time $t_{0}$ leaves the future states of the system unspecified \cite{Bergmann,BergmannLectures}, even given the past.  This problem is especially familiar to philosophers of physics now due to the hole argument \cite{HoleStory}.   
  Even though there is a flat metric with a complete set of commuting Killing vector fields that define \emph{preferred} coordinate systems in the Klein-Gordon field with the Rosen-Sorkin trick, Earman demands that observables be invariant under \emph{all} coordinate transformations.  Some of these transformations start with a set of preferred (Cartesian) coordinates and conclude with non-preferred (non-Cartesian) coordinates; some others start with non-preferred (Cartesian) coordinates and conclude with preferred (Cartesian) coordinates; others do neither.  Earman is strangely silent about the role of Killing vector fields for the Klein-Gordon scalar field theory with the Rosen-Sorkin Lagrange multiplier trick.  By contrast, Bergmann  would reject Earman's criterion for observability, because given a specification of the past of the system and due attention to the invariant Killing vector fields, the future coordinatization of space-time already is determined in a preferred way up to a global  Poincar\'{e} transformation at the worst. That  Bergmann would say this follows from his distinction between weak and strong general covariance and his dismissive attitude toward the former \cite{BergmannLectures,StachelGC}.  I will quote Bergmann at length from a work that is not readily available:
\begin{quote}
Look, for a moment, at the familiar Lagrange equations of classical mechanics.  Let us start with some given coordinate coordinate system (which we may assume has a physical---i.e., operational---meaning, quite independently of the dynamics of our mechanical system).  We write down Lagrange's equations.  If we will, we may pass to a second coordinate system---bearing always a known relation to the first---and again write the Lagrange equations.  We rejoice that, in a certain well-defined sense, these new equations exhibit the same structure as the old...and announce that Lagrange's equations are ``covariant.''  This is convenient: \emph{it is  not profound.}  This ``weak'' form of covariance affords us an efficient way to describe the \underline{same} physical motion in any coordinate system that pleases us.  We are not disturbed if in different coordinate systems our equations of motion \underline{look} different...for we know they are different descriptions of the same orbit: we know how to pass from one description to another, and the operational relation of each to our world.  Our theorem [that covariance disallows predictions], if it were to be interpreted in this context, would be scarcely worth stating. 

The fundamentally trivial nature of this ``weak covariance'' derives from the rigidity of the classical metric.  When this is relaxed, we arrive at the deeper notion  of ``strong covariance''---the proper context for understanding our theorem.  For in general relativity it is one's task to \underline{calculate the metric}...as a dynamical variable.  We can take one coordinate system or another for this job, but all that we can know is the relation of one frame to the other: we do not know the relation of \underline{either} to the world.  ``Strong covariance'', therefore, contains not only a reference to the structural similarity of an equation and its transform; it implies as well that \emph{one frame is as good a starting point as another}'' \cite[p. 11]{BergmannLectures}. (italics added, ellipses and underlining in original) \end{quote} 
In discussing the Klein-Gordon theory with the Lagrange multiplier trick, Earman ignores the fact that the flat metric's Killing vector fields pick out preferred coordinates.  They are preferred only up to a global Poincar\'{e} transformation, but the global nature of the symmetry means that the choice is made only once and also is encoded in the coordinate expression of the initial data from which one wants to predict the future  
\cite{RovelliObservable,PonsObservable}.  Thus the predictability problem that explicitly motivated Bergmann's definition of observability  does not arise in STR, even if a weakly covariant formulation like the Klein-Gordon scalar field with the Lagrange multiplier field is used.  Thus Bergmann would reject Earman's appeal to observability in the technical sense in application to a scalar field theory in Minkowski space-time with the Rosen-Sorkin Lagrange multiplier trick.   Once Earman's doubtful observability criterion is diagnosed, his motivation for regarding the two scalar field formulations as distinct theories is removed.
Given that Sorkin's  scalar field formulation  can be gauge-fixed into the ordinary scalar field formulation as far as the scalar field's evolution is concerned,  there is no evident reason to regard the two as distinct theories.  Giulini's suggestion that the evolution of the Lagrange multiplier field gives distinct observable content to the Sorkin formulation \cite{GiuliniAbsolute} is mistaken (unless one imagines observables to be built out of the Lagrange multiplier field!), because the Lagrange multiplier does not appear in the scalar field equation (and so is irrelevant in Anderson's sense \cite{Anderson,FriedmanJones}).   
Thus one can leave Pandora's box closed by retaining the usual practice of identifying theory-candidates related by gauge transformations and/or gauge fixings.  Such a result is  reassuring regarding physicists' practice in quantum field theory.  It might be interesting to \emph{test} a criterion for observability by demanding that it yield essentially the same observables for the Lagrange multiplier  scalar field formulation as  for the ordinary Klein-Gordon scalar field formulation. The presence of a complete commuting set of Killing vector fields in every model---in short, preferred coordinate systems---should not be ignored.  Cartesian coordinates, though not necessary for prediction, are certainly adequate \cite{BradingBrown}.

Earman goes on to consider unimodular general relativity \cite{EarmanCovariance}. 
 The Bergmann-inspired criticism of the application of the concept of observables to theories that surely are not substantively generally covariant indicates that more work is needed before relying much on differing lists of observables as Earman does.  He contemplates the idea that the use of clock fields ``does not satisfy the spirit of general covariance'' and notes that this ``complaint invites us to produce an even stronger version of the requirement of general covariance than the one proposed here.  I would decline the invitation.'' \cite[p. 462]{EarmanCovariance} By contrast, I accept the invitation and note that Earman's example of clock fields is precisely suited to motivate using the absence of clock fields as the core of the test for substantive general covariance.


The inter-convertibility of clock fields with nonvariational fields addresses unimodular GTR with clock fields rather nicely, because then, for example, the formulation of unimodular gravity with clock fields is readily converted to a formulation with a nonvariational background volume element, which by Earman's standards violates substantive general covariance.  Given that clock fields and nonvariational fields are inter-convertible, a criterion for substantive general covariance that bans nonvariational fields can only succeed if it also bans clock fields.  Thus one might analyze  substantive general covariance as follows.
\begin{quote}  {\bf Definition.}  A field theory is substantively generally covariant just in case it is formally generally covariant (in the sense of admitting at least arbitrary infinitesimal coordinate transformations and some finite transformations near the Lorentz group), lacks irrelevant fields (in the sense of James Anderson \cite{Anderson,FriedmanJones}), lacks nonvariational fields and lacks  clock fields.  
\end{quote}

To my knowledge this criterion performs as expected, extensionally, with perhaps some ambiguity if the gauge $X^M = x^{\mu}$ is not permitted, as in massive gravity with gauge freedom \cite{MassiveGravity1}; gauge freedom is required to avoid causality problems. The hesitation in requiring the admissibility of arbitrary finite coordinate transformations is intended to secure the general covariance of a spinor formalism \cite{OPspinor,IshamSalamStrathdee,ChoFreund,BilyalovSpinors} that is more parsimonious and perhaps defined with less demanding assumptions than is the usual spinor formalism with an orthonormal tetrad.
This  criterion of lacking clock fields, however, clearly is at heart much like Einstein's original claim that GTR was generally covariant in the sense of admitting expression using arbitrary coordinates.  To the Kretschmann objection that any theory could be so formulated---one can imagine a counterfactual history in which Kretschmann produces the example of parametrized field theories in Minkowski space-time in response to Einstein's proposal---one could reply on Einstein's behalf that Kretschmann's general covariance was artificial because it introduced clock fields as a very transparent disguise for preferred coordinates, whereas GTR does no such thing.  Thus there is some hope for distinguishing the real general covariance of GTR from the artificial sort.

 It is perhaps ironic that so old-fashioned a criterion as the absence of preferred coordinates, with some modern polish, works better than the  more sophisticated Anderson-Friedman project and appears to give the desired answers, at least if one wishes to retain the claim that GTR is substantively generally covariant. (Sean Carroll's view has some similar elements \cite{CarrollSpacetimeGeometry}, as does that of Goldhaber and Nieto \cite{GoldhaberNieto2009}.) The absence of clock fields and the absence of fields not varied in the action principle seem to be not merely coextensive, but very simply related conceptually in light of the inter-convertibility of clock fields and nonvariational fields.  The claim that there is nothing much interesting about coordinates is overstated; while it is true for GTR, it is not true in general, and this very fact sheds light on GTR.  Of course clock fields $X^A$ have no exemption from the problem that some manifolds cannot be covered with a single coordinate chart; thus one might need to piece together generalized clock fields out of multiple charts.   

If there is in fact   extensional equivalence between the purely variational definition and the parametrized definition of substantive general covariance, there is no need to choose between them if one merely wants to classify a given theory as substantively generally covariant or not.  Elsewhere  I suggested  that the variational formulation gives the right answer for perhaps  the wrong reason in the case of GTR \cite{PittsPhilDiss}.  Does a similar worry arise for clock fields?  While I admit surprise 
 that the sophisticated Anderson-Friedman analysis fails and the simple coordinate analysis of the early Einstein essentially succeeds, presently I see no objection to taking the absence of clock fields (and irrelevant fields and nonvariational fields) as an adequate analysis of general covariance, if there is one.  Once one bans nonvariational fields and irrelevant fields,  formal general covariance without clock fields apparently does imply substantive general covariance.

The parametrized form of a certain massive variant of GTR \cite{FMS,Schmelzer,Arkani,MassiveGravity1} below will further illustrate the wisdom of using clock fields as a test for general covariance.    Installing artificial gauge freedom by parametrization at least formally restores many features of GTR that one might associate with substantive general covariance:  gauge freedom,  point individuation questions such as appear in the hole argument, the absence of non-variational fields in the Lagrangian density, a vanishing Hamiltonian apart from boundary terms, \emph{etc.}  It is by no means clear that massive variants of GTR are physically acceptable, it should be noticed, due to worries about stability.  This question will be discussed a bit more below.  However, it is not crucial that massive gravities be physically acceptable in order to make the conceptual points discussed here.

 As a further example of the power of clock fields to simulate  substantive general covariance in a deceptively plausible way, one might  consider GTR as formulated with an  ADM split of space-time into space and time \cite{MTW}, which introduces a temporal foliation.  One can add the time foliation as a new scalar  field $T$, varied in the action principle, at no cost, because the resulting Euler-Lagrange equation is entailed by the others.  One now has the ingredients for building lots of \emph{new} theories using 3-dimensional entities, which in turn are defined using 4-dimensional tensors and $T$.  Of course most of these theories, unlike GTR, will make ineliminable reference to $T$  and contain absolute simultaneity observably.  But the theories' Lagrangians are built entirely using 4-dimensional tensors and the `scalar' clock field $T$.  Clearly introducing the clock field $T$ into the theory evacuated the 4-dimensional symmetry of content while preserving it formally.  Given that clock fields are such outstanding tools for subverting otherwise promising criteria for substantive general covariance, it seems advisable (assuming that one has already converted all nonvariational fields into clock fields) to understand substantive general covariance in terms of the absence of clock fields.  




\section{Artificial Gauge Freedom}

It is well known that many important physical theories have considerable descriptive redundancy.
The  gauge theories for the four fundamental forces---Maxwell's electromagnetism, the Yang-Mills theories for the weak and strong nuclear forces, and GTR---display this descriptive redundancy.  Though gauge freedom arguably has a deep philosophical significance of some kind, gauge freedom poses something of a challenge technically.  In some contexts, especially Maxwell's electromagnetism or its quantum successor, it can be useful to make a conventional choice of a specific gauge, make calculations in a convenient fashion, and then show that the results did not depend on that specific gauge choice \cite[p. 345]{WeinbergQFT1}.  
Unfortunately this simple procedure tends to fail for Yang-Mills theories due to the Gribov ambiguity \cite{Sundermeyer,Kaku,Guay}; this failure can be relevant when nonperturbative effects matter. 
A more elegant, but more abstract and technically difficult procedure, is to work with a space of physically distinct configurations by taking gauge equivalence classes; the resulting reduced phase space (in a Hamiltonian formalism) or other reduction to the ``true degrees of freedom'' might be difficult to find or use explicitly, however.  These sorts of procedures take gauge freedom to be an obstacle to overcome.  Thus Freund, Maheshwari and Schonberg (FMS) discussed how the absence of gauge freedom in their massive variant of GTR made it  easier to quantize than GTR \cite{FMS}.  One can take the second-class constraints as identities for eliminating unnecessary field variables \cite{Sundermeyer,WeinbergQFT1}.
In terms of the Dirac-Bergmann constrained dynamics \cite{Baaklini,Marzban,PittsQG05}, massive GTR has only second-class constraints and thus makes the true degrees of freedom available almost immediately in terms of Dirac brackets (which  generalize Poisson brackets), at least in principle. 
However, for many theories  there are technical challenges involved in working with only the true degrees of freedom \cite{HenneauxTeitelboim,BanerjeeGhosh,ParkPark}, such as  that (i) Dirac brackets tend to be nonlocal and field-dependent, 
(ii) for quantization, finding a suitable factor ordering (recalling that $qp-pq \neq 0$ in quantum mechanics) is often difficult or perhaps impossible, and (iii)  finding a complete set of variables to form canonical pairs is often difficult or perhaps impossible.

The pragmatic challenges in dealing with Dirac brackets and  the progress in handling gauge theories have led to a change of viewpoint.  More recently the view has appeared that gauge freedom is an asset and not so much a liability. Thus technologies have been developed to take theories with constraints but no gauge freedom (of which the FMS and other massive gravities are examples \cite{OP,OPMassive2,FMS}, as are the Proca massive electromagnetisms) and \emph{install} gauge freedom artificially by adding extra fields and extra symmetries ensuring that the extra fields make no empirical difference.  While the \emph{ad hoc} Stueckelberg trick developed gradually in the middle of the 20th century \cite{Ruegg}, 
 and a paper from the 1970s left the name ``Wess-Zumino fields'' for certain purposes \cite{WessZumino,NetoRemove}, the subject of installing artificial gauge freedom reached a mature form in the 1980s   \cite{FaddeevShatashvili,BF86,BF87,BFF89,BT91} and now bears the names of Batalin, Fradkin and Tyutin, or BFT for brevity.   
 This sort of technique has been applied to both Proca's massive electromagnetism, 
  and for a nontrivial test case, massive Yang-Mills theories (notwithstanding the apparently defective character of most such theories under quantization) \cite{KunimasaGoto,Slavnov,Shizuya1,Shizuya2,GrosseKnetter,BanerjeeBN,YMembed,KimParkYoon,Ruegg}.   
I anticipate that application of the BFT  procedure  to massive gravities with  $6 \infty^3$ degree of freedom \cite{OP,OPMassive2,FMS} would reproduce the result of parametrization with clock fields \cite{Schmelzer,Arkani,MassiveGravity1}, apart from some issues to be discussed below.  
Such techniques are usually formulated in a Hamiltonian context, where the well-developed Dirac-Bergmann constrained dynamics formalism \cite{Sundermeyer,HenneauxTeitelboim,EarmanGauge,EarmanOde} allows one to speak of BFT ``conversion'' of second-class constraints into first-class constraints.

 There has also been a bit of work in a Lagrangian context   \cite{ParkPark,KimParkYoonEW}.  The Lagrangian formalism provides considerable help in guessing correct answers.  Given the conceptual distinction between the context of discovery and the context of justification \cite{ReichenbachEP}, one need not be ashamed of a guess-and-check procedure when systematic treatment is difficult.    Here there are two advantages.  First, the Lagrangian formalism introduces only half as many new fields as the Hamiltonian formalism:  new fields but not new conjugate momenta.
Having no new momenta can   significantly reduce the number of candidate terms in the power series expansion of the new piece of the Lagrangian as opposed to the new piece of the Hamiltonian.  Second and perhaps even more importantly, the manifest Lorentz covariance of the Lagrangian formulation of relativistic theories still further reduces the number of candidate terms.  In general I will limit attention to Hamiltonian formulations of theories, but I prefer Hamiltonians related by Legendre transformations to Lagrangians with the usual symmetries (gauge invariance, Lorentz covariance, \emph{etc.}) manifest when possible.  (GTR might be an exception, but recently the fortunes of the Hamiltonian treatment of Einstein's $\Gamma\Gamma$ Lagrangian have drastically improved \cite{Kiriushcheva}.)
Below I will make use of a Lagrangian `lucky guess' in installing artificial gauge freedom for FMS massive gravity. The availability of a Lagrangian constraint stabilization algorithm analogous to the Dirac-Bergmann Hamiltonian algorithm \cite{ShepleyEvolutionary} implies that at least some cases can be treated directly via Lagrangian means.



The typical BFT procedure goes along these lines:  start with a Hamiltonian theory (formulation) with $m$ second-class constraints and no first-class constraints.  Introduce $\frac{m}{2}$ pairs of new coordinates $\theta$ that are canonically conjugate (in a slightly generalized sense).  Take the original (second-class) constraints and Hamiltonian as    the zeroth order terms in a series expansion involving the new fields $\theta$.  Choose the coefficients in the series  to make all constraints first-class and make them have vanishing Poisson brackets with the Hamiltonian  \cite{BF86,BF87,BFF89,BT91,VytheeswaranAnnals,VytheeswaranHidden}.  Given the series expansion, the original formulation is recovered in the limit that the new fields vanish.
One can count the degrees of freedom to show that the new formulation has the same number as the old.  Given $l$ configuration variables (and hence $2l$ phase space variables), $f$ first-class constraints, and $s$ second-class constraints in a theory formulation, there are $\frac{2l -2f -s}{2}$ degrees of freedom \cite{HenneauxTeitelboim}.  For the original formulation, there are $\frac{2l -2\cdot 0 -m}{2}= \frac{2l  -m}{2}$ degrees of freedom. For the new formulation, there are $\frac{(2l+m) -2m -0}{2} = \frac{2l  -m}{2}$ degrees of freedom, the same as for the original.  
  The new formulation with first-class constraints presumably allows gauge transformations to achieve $\theta=0,$ so the new formulation can be gauge-fixed into the old one by choosing $\theta=0$ (unless some physical principle would be violated thereby, such as causality as in massive gravity \cite{MassiveGravity1}).

\section{Modified  Batalin-Fradkin-Tyutin Procedure Gives Stueckelberg Formulation Directly}

The most compact and perspicuous way to begin a technical discussion of a classical field theory is to exhibit its Lagrangian density, a function of some fields and their derivatives, such that the space-time integral of the Lagrangian density $\mathcal{L}$, the ``action'' $S$ of the theory, satisfies the principle of least (or perhaps merely stationary) action.  
  The source-free Maxwell field equations (in manifestly Lorentz-covariant form) follow from a Lagrangian density of the form \begin{equation} \mathcal{L}= -\frac{1}{4}  F_{\mu\nu}  F^{\mu\nu}, \end{equation} where the indices are moved using the Lorentz metric $diag(-1,1,1,1)$ and $F_{\mu\nu}=_{def} \partial_{\mu} A_{\nu}- \partial_{\nu} A_{\mu}$ is the electromagnetic field strength.   For Maxwell's theory, the vector potential $A_{\mu}$ admits the gauge transformation $$ A_{\mu} \rightarrow A_{\mu} + \partial_{\mu} \phi$$ for an arbitrary function $\phi;$ this transformation makes no observable difference.  This Lagrangian density is manifestly gauge invariant, because it is built from the gauge-invariant field strength only. For the massive Proca electromagnetisms, the Lagrangian density  is \begin{equation}  \mathcal{L}_p = -\frac{1}{4} F_{\mu\nu}F^{\mu\nu} - \frac{m^2}{2} A_\mu A^\mu. \end{equation}  Evidently the $A^2$ term breaks the gauge symmetry in the massive case.  Whereas Maxwell's theory has $2$ degrees of freedom at each spatial point (written as $2\infty^3$ degrees of freedom), Proca's theories have $3\infty^3$ degrees of freedom.\footnote{The reader will observe the importance of distinguishing  $2\infty^3$ from  $3\infty^3$, notwithstanding rules for Cantorian transfinite arithmetic \cite{Moore}.  The lesson seems to be that physical theories involve continuity properties of sets from which cardinality abstracts.  Evidently cardinality does not exhaust the useful notions of ``same size'' or counting for infinite collections.
}  The extra degree of freedom (at each point), however, is  weakly coupled for small photon masses and so is not readily noticed experimentally.  The treatment of the two theories (or theory types) using the Dirac-Bergmann constrained dynamics formalism is straightforward \cite{Sundermeyer}.  The approximate empirical equivalence between Maxwell's theory and Proca's theories for small enough photon masses is preserved under quantization:  massive quantum electrodynamics (QED)  approximates the standard massless QED arbitrarily well \cite{BelinfanteProca,Glauber,BassSchroedinger,StueckelbergMasslessLimit,BoulwareGilbert,BoulwareYM,GoldhaberNieto,SlavnovFaddeev,DeserMass,Slavnov,Shizuya1,Ruegg,GoldhaberNieto2009}.  
In thermal contexts, where one might expect the third degree of freedom to be relevant, it decouples in the massless limit, so that the time to reach equilibrium is inversely related to the photon mass.  As the mass goes to zero, a system takes forever to reach equilibrium; hence equilibrium thermodynamic quantities based on three field degrees of freedom are physically irrelevant and unobservable \cite{BassSchroedinger,GoldhaberNieto}.  
It follows that in a world with electromagnetism as the only force, it would be impossible for finite beings
to rule out all of the massive electromagnetic theories empirically, and thus impossible to determine empirically whether gauge freedom was a fundamental feature of the electromagnetic laws.  Here I am forgetting about the Stueckelberg formulation to be discussed shortly, which shows all the more strongly that gauge freedom \emph{per se} is not even distinctive of massless electromagnetism, unless one bans certain extra gauge compensation fields.

Given the starring role playing by Maxwell's electromagnetism, Yang-Mills theory, and Einstein's gravity in contemporary physics, and consequently the non-negligible importance of their massive cousins (whether those theories are physically viable or not),  it seems reasonable to seek a procedure for converting non-gauge formulations into gauge formulations that is as convenient as possible for massive electromagnetism, massive Yang-Mills theory, and massive GTR.  In the electromagnetic and gravitational cases, it turns out that the gauge formulations were found in an \emph{ad hoc} way without the need for such elaborate conversion algorithms as have appeared in recent decades.  In the electromagnetic case, the answer is Stueckelberg's trick,  which adds the gradient of a new scalar field to the vector field $A_{\mu}$ in the mass term; adding such a term in the kinetic term $ - \frac{1}{4} F_{\mu\nu}F^{\mu\nu}$ would make no difference because such a term, like a gauge transformation, has no effect on $F_{\mu\nu}=\partial_\mu A_\nu - \partial_\nu A_\mu.$  In the gravitational case, the answer is parametrization,  whereby preferred coordinates are turned into new fields varied in the action principle.  One hopes that parametrization reproduces the outcome from a BFT-type procedure, though a full direct nonlinear calculation seems never to have been done.  (In fact one would expect the relationship to be somewhat indirect, for reasons to be discussed shortly.) For the Yang-Mills works cited above, it turns out that the answer is too complicated to guess easily, but the problem  yielded to systematic treatment.  


The usual BFT procedure makes no use of a ``polarization'' of the phase space into configuration and momentum variables, a fact that adds some generality but leaves the results a bit more complicated than is necessary for use in some paradigmatic cases, such as Proca or massive Yang-Mills theories.  Whereas one often sees the BFT procedure yield the Stueckelberg formulation of massive electromagnetism only after a rather elaborate calculation involving  a canonical transformation to implement a change of variables,  path integration, and  dropping a boundary term  \cite{Banerjee,VytheeswaranCompare,BanerjeeGhosh}, I will now discuss a modified BFT formalism that  directly recovers the Stueckelberg version of massive electromagnetism. The modification involves making a distinction between coordinates and momenta in a fundamental way,
for both the original field variables and the additional ones added to install gauge freedom.  While some generality might be lost thereby, 
there is a formal simplification that might be important in applications, such as gravitation.  It was a considerable achievement in 1958 when it was found how to trivialize the form of the primary constraints in canonical GTR \cite{DiracHamGR,AndersonPrimary}, 
though it has been shown recently that perseverance with the manifestly Lorentz-covariant Einstein $\Gamma\Gamma$ Lagrangian quadratic in the Christoffel symbols has some benefits \cite{Kiriushcheva}. The modification made herein to the BFT procedure respects this achievement of trivializing the primary constraints by leaving the primary constraints alone.  It also distinguishes among the new fields a set of new coordinates and new momenta.  The result is  more Lagrangian-friendly than is the usual BFT procedure, as well as pedagogically simpler.  It is optimized for convenience in paradigm cases, but might suffer in generality, however.

From a Lagrangian point of view, one wants to recover the second-class formulation from the first class formulation by imposing the vanishing of new coordinates  (such as the Stueckelberg scalar) and their \emph{time derivatives} (`velocities'), rather than the vanishing of new fields and momenta.  The mathematical details will make this discussion clearer; they also serve as a review of relevant parts of the Dirac-Bergmann constrained dynamics formalism \cite{Sundermeyer}.  For the massive Proca electromagnetism, the Lagrangian density is  
$$ 
\mathcal{L}_p = -\frac{1}{4} F_{\mu\nu}F^{\mu\nu} - \frac{m^2}{2} A_\mu A^\mu.$$ 
Here as elsewhere I used the $-+++$ signature, letting Greek indices run from 0 to 3 and  Latin indices from 1 to 3.
  The primary constraint (at each point) is $$\pi^0 = \frac {\partial \mathcal{L}_p }{\partial A_{0,0} } = 0.$$ Performing the Legendre transformation to get the canonical Hamiltonian density (not a redundant term in the constrained dynamics literature \cite{Sundermeyer}) gives
$$ \mathcal{H}_{pc} = \pi^\alpha A_{\alpha,0} - \mathcal{L}_p = \frac{1}{2}(\pi^a)^2 + \pi^a A_{0,a} +\frac{1}{4} F_{ij}F_{ij} + \frac{m^2}{2} (A_i)^2  - \frac{m^2}{2} (A_0)^2.$$  Using the canonical Hamiltonian $ \int d^3 x \mathcal{H}_{pc} $ (or the primary Hamiltonian---it does not matter) to find the time evolution of $\pi^0 $ and to enforce its continued vanishing gives
$$\{ \pi^0(y),  \int d^3 x \mathcal{H}_{pc} (x) \} = \pi^a ,_a (y) + m^2 A_0 (y) \approx 0.$$ 
Thus the secondary constraint is $\pi^a ,_a  + m^2 A_0 $ everywhere.  Demanding that the secondary constraint be preserved by the time evolution requires the use of the primary Hamiltonian $H_{pp} = H_{pc} +\int d^3 x v(x) \pi^0 (x),$ where $v$ is a new Lagrange multiplier.  
Preserving the secondary constraint (with the help of an arbitrary test function to smear the Dirac delta functions from the fields' Poisson brackets as needed) gives 
$$\{ \pi^a ,_a (y) + m^2 A_0 (y)  ,  {H}_{pp} \} = - m^2 A_i ,_i + m^2 v \approx 0,$$ fixing $v$ and leaving no arbitrariness in the evolution of the system.  The Poisson brackets of the constraints among themselves are
$$ \{ \pi^0 (x), \pi^0 (y) \} =0,$$  $$\{ \pi^0 (x), \pi^a ,_a + m^2 A_0 (y) \} = -m^2 \delta(x,y), $$ and $$\{ \pi^a ,_a + m^2 A_0 (x), \pi^i ,_i + m^2 A_0 (y) \}=0.$$  
The vanishing Poisson brackets here vanish without the use of the constraints themselves.  The matrix of Poisson brackets of constraints has non-vanishing determinant, so the theory is indeed second-class as advertised.  

How does the constrained dynamics treatment of the Stueckelberg gauge formulation of massive electromagnetism differ?  The Lagrangian density is 
$$ 
\mathcal{L}_{s} = -\frac{1}{4} F_{\mu\nu}F^{\mu\nu} - \frac{m^2}{2} (A_\mu + \partial_\mu \phi) (A^\mu+ \partial^\mu \phi).$$  As with clock fields, it does not matter if the Stueckelberg scalar $\phi$ is varied or not. The reason is that the gauge transformation formula for the Stueckelberg field $\phi$ is algebraic in the gauge parameters (as is also the case for massive Yang-Mills \cite{YMembed} and for massive gravity \cite{Schmelzer,Arkani,MassiveGravity1}).   If $\phi$ is not varied, then its erstwhile equation of motion will still follow from the Euler-Lagrange equation for $A_{\mu},$ either using the generalized Bianchi identity or by simply taking the divergence.  Turning to the Hamiltonian formalism, the primary constraint is $\pi^0 = \frac {\partial \mathcal{L}_{s} }{\partial A_{0,0} } = 0,$ as before; the momenta for $A_i$ are also unchanged.  The new momentum for the new field $\phi$ is $$P = \frac {\partial \mathcal{L}_{s} }{\partial \phi,_0 } = m^2 A_0 + m^2 \phi,_0, $$ so the vanishing of the new momentum (part of the BFT boundary condition) and the vanishing of the new field's time derivative (part of the new boundary condition proposed here) are in general incompatible.   Performing the Legendre transformation to get the canonical Hamiltonian density  gives
$$ \mathcal{H}_{sc} =  \frac{1}{2}(\pi^a)^2 + \pi^a A_{0,a} +\frac{1}{4} F_{ij}F_{ij} + \frac{m^2}{2} (A_i)^2 + \frac{P^2}{2m^2} -A_0 P +m^2 A_i \phi,_i + \frac{m^2}{2}(\phi,_i)^2.$$  
Preserving the primary constraint yields the secondary constraint $$\pi^a ,_a + P \approx 0.$$  Preserving the secondary constraint gives neither a fixed Lagrange multiplier $v$ nor a tertiary constraint.  The primary constraint, being unchanged, still Poisson-commutes with itself.  The new secondary constraint also Poisson-commutes with itself.  Most importantly, 
$\{ \pi^0 (x), \pi^a ,_a + P(y) \} = 0,$ so the Stueckelberg formulation is first class, as advertised.  This is the same Poisson bracket algebra as in the Maxwell theory. One advantage of the Stueckelberg formulation of massive electromagnetism over the Proca formulation is the ease of taking the massless limit \cite{Zinoviev}; the gauge freedom remains, while the extra degree of freedom decouples as $m \rightarrow 0.$

  Comparing the secondary constraints of the two formulations is illuminating.  For the  Proca formulation, the secondary constraint is $\pi^a ,_a  + m^2 A_0, $ whereas the secondary constraint for the Stueckelberg formulation is $\pi^a ,_a + P .$ If one pretends that the Stueckelberg formulation has been produced by the usual BFT procedure with the boundary condition of getting the old formulation when the new fields all vanish, then one should expect the new constraints to reduce to the old ones once the new fields $\phi$ and $P$ are set to $0.$  The primary constraint, being unchanged, satisfies this expectation, but the secondary constraint does not:  $\pi^a ,_a + P $ becomes $\pi^a ,_a + 0 \neq   \pi^a ,_a  + m^2 A_0. $ This result is disappointing, if one hoped that the BFT procedure would reproduce conveniently  the Stueckelberg formulation.  On the other hand, the relation $P =  m^2 A_0 + m^2 \phi,_0$ already told us that the vanishing of the new momentum $P$ generally is inconsistent with the vanishing of the new velocity $\phi,_0,$ as was noted above. Perhaps one could circumvent this problem with a canonical transformation,  
but why work so hard?  Another response to this difficulty would be to replace the BFT boundary condition that new constraints reduce to the old ones when the new fields (here $\phi$ and $P$) vanish, with the condition that the new constraints reduce to the old ones when \emph{those new fields that appear in the Lagrangian density} vanish (here $\phi$) and the appropriate relation between the new momenta and the new velocities holds.  The relation between the new momenta and the new velocities arises, from the Hamiltonian angle, from the equation of motion $\phi,_0 = \frac{ \delta H_s}{\delta P}$:  the functional derivative of the new Hamiltonian with respect to the new momenta is made to vanish.    Thus I propose imposing $\phi=0 $ and $\frac{ \delta H_s}{\delta P}=0,$ rather than $\phi=0$ and $P=0,$ as the means to recover the Proca formulation by gauge-fixing the Stueckelberg formulation. Imposing my conditions makes the Stueckelberg secondary constraint and Hamiltonian reduce to the Proca secondary constraint and Hamiltonian, as desired.  One can also consider whether a set of extra conditions suffices to fix the gauge by forming, along with the primary and secondary constraints, a matrix of Poisson brackets with non-vanishing determinant \cite{Sundermeyer}.  It is easy to show that the conditions  $\phi=0$ and $P=0$ fail to gauge-fix the Stueckelberg formulation  because the resulting four conditions $\pi^0,$ $\pi^a ,_a + P, $  $\phi=0$ and $P=0$ fail to form a matrix of Poisson brackets with non-vanishing determinant.  By contrast, my conditions  $\phi=0$ and $\frac{ \delta H_s}{\delta P}=0$  do fix the gauge.  In general my proposal involves making a fundamental distinction among the new fields between new coordinates and new momenta, unlike the usual BFT procedure.

One might not like the fact that the  BFT procedure sometimes changes  the primary constraints.  In the Stueckelberg formulation (not resulting from the BFT procedure), it was clear above that all the old momenta are unchanged relative to the Proca formulation, and indeed relative to the massless Maxwell case.  Given that primary constraint $\pi^0$ has vanishing Poisson bracket with itself, all the non-vanishing Poisson brackets that make the Proca formulation second-class in Hamiltonian form evidently involve the secondary constraint $\pi^a ,_a  + m^2 A_0.$  Clearly it is no fault of  the \emph{primary} constraints that the Proca theory has second-class constraints, then.  So why modify the primary constraints, if one can avoid it?  Leaving the primary constraints unchanged should also remove some of the arbitrariness inherent in the BFT procedure.  If the primary constraints are left unchanged, then the change in the Hamiltonian entails a definite change in the secondary (and higher order, if any) constraints.


One can  
 derive, honestly and simply (with no dropping of boundary terms, field redefinitions through canonical transformations, or functional integration), the Stueckelberg formulation starting with the Proca formulation.
Retaining the old primary constraints and introducing one new field and corresponding new momentum, one makes an unknown change in the Hamiltonian  $\Delta H,$  and hence a corresponding change in the secondary constraints.  Then one  seeks the following:
the dynamical preservation of the new secondary constraints by the new Hamiltonian, 
the first-class nature of all constraints in the new formulation, 
and the boundary condition that the vanishing of the new coordinate only (not the new momentum) and the vanishing of the  derivative of the new Hamiltonian with respect to the new momentum gives the Proca Hamiltonian. The resulting Poisson bracket algebra is the same as in Maxwell's theory.  Making an inverse Legendre transformation from the gauged Hamiltonian formulation yields precisely the usual Stueckelberg Lagrangian. 
It would be of interest to apply the modified BFT procedure to Yang-Mills theories as well.





While the treatment in this section has been classical, the major physical (as opposed to philosophical) benefits of installing artificial gauge freedom to convert the Proca formulation into the Stueckelberg formulation lie in quantum field theory. As noted in the papers cited above, the propagators for the Proca formulation have bad high-energy behavior, but the Stueckelberg formulation avoids that difficulty.  Whether such an issue would arise in some future formulation of quantum field theory that avoids some of the crutches that seem  difficult to do without at present, if there is one, would be interesting to know.  The philosophical benefits include providing a tractable yet interesting  and physically realistic illustration of a generalized Kretschmann objection.  If Proca's non-gauge electromagnetism can be reformulated into Stueckelberg's gauge version, the significance of the gauge freedom of Maxwell's electromagnetism is less clear.  Hints have been given for answering this objection also.

	
\section{Gauge Freedom in Massive Gravity: An External Symmetry}

Whereas the previous section considered internal symmetries, for which each point lives a life by itself, there are other symmetries in which gauge transformations link what goes on at different space-time points.  The most famous of these are the ``external'' space-time diffeomorphisms. 
Less famous but quite important are the transformations of supersymmetry, in particular of supergravity \cite{vanNReports}, in which internal and external symmetries are combined. 
 Also of interest are the  gauge transformations of Einstein's equations on a background, which are a symmetry distinct from diffeomorphisms but sharing some of their mathematical features \cite{Grishchuk,NullCones1,PetrovChapter}.  Thus ``internal'' and ``external'' symmetries are neither cleanly mutually exclusive (being in some cases mixed together\footnote{I thank a referee for pointing out that global bundle properties provide another context where the division between internal and external symmetries can be problematic.}) nor exhaustive. 

 There is a large and growing literature on massive gravities \cite{OP,FMS,DeserMass,Vainshtein,Visser,Karch,Vainshtein2,MassiveGravity1,Zinoviev}, much of it addressing whether they approach (local) empirical equivalence with Einstein's equations in the massless limit and are theoretically healthy, even at the classical level. 
A widely held view since the early 1970s poses a dilemma  \cite{DeserMass,TyutinMass} asserting  that massive gravities either have $5\infty^3$ degrees of freedom (spin 2) and do not agree with Einstein's equations in the massless limit due to the van Dam-Veltman-Zakharov discontinuity \cite{vDVmass1,vDVmass2,Zakharov}, or they have $6 \infty^3$ degrees of freedom (spin 2 and spin 0) and agree empirically with Einstein's theory in the massless limit (at least classically), but are theoretically unhealthy and physically unstable because the spin 0 field has negative kinetic energy.  This view has been challenged in recent years \cite{Visser,GrishchukMass}.  Whereas the Proca theory is the unique local linear massive variant of Maxwell's electromagnetism, the most famous massive gravity with $6 \infty^3$ degrees of freedom, the FMS massive gravity \cite{FMS,DeserMass,LogunovBook}, is just one member (albeit the best in some respects) of a 2-parameter family of massive theories of gravity \cite{OP,OPMassive2}, all of which satisfy universal coupling \cite{MassiveGravity1}. Whether massive gravities are viable even at the classical level remains a matter of debate in the physics literature. Evidently intuitive expectations about the ease of constructing approximately empirically equivalent theories to GTR are threatened by devils (or rather, ghosts, implying negative energies or negative probabilities) in the details, but the arguments are not yet conclusive. Standard theorems about the disastrous character of negative energy degrees of freedom do not apply if the Hamiltonian is not separable into the sum of a function of the momenta and and a function of the coordinates  \cite{Morrison}. 
Massive variants of GTR do not display that simple form \cite{MassiveGravity1}. Even some Hamiltonian theories that are separable and that have negative energy degrees of freedom can be stable in the absence of resonances \cite[p. 81]{MeyerHallOffin}.  Plasma physicists had considerable experience making detailed investigation of systems with negative energy waves.   The presence of absence of resonances and the existence or nonexistence of a reference frame in which all members of a three-wave resonance have the same energy sign are key factors \cite{DavidsonPlasma,WeilandWilhelmsson}.   
While instability is indeed a serious possibility, it is not a foregone conclusion in every context.   Thus the situation at the classical level is not entirely clear.  Nothing requires the spin 0 field to have the same mass as the spin 2 \cite{OP}, so there is room to tweak the relativistic dispersion relations. Such numerical simulations of massive gravity (with $6 \infty^3$ degrees of freedom) as have been done have failed to display the expected instability \cite{GrishchukMass}.

A further question is what becomes of such massive theories under quantization.  While one might worry that they suffer from negative energy states or negative probabilities, that outcome might not be a foregone conclusion.  
One possibility worthy of exploration is whether the methods of PT-symmetric quantization can help.  PT-symmetric quantization has  exorcised the vicious ghosts thought to inhabit some theories according to more traditional analyses \cite{Mostafazadeh,Bender,BenderMannheim}, though the resulting theories sometimes have surprising phenomenology.  Might PT-symmetric be helpful for massive gravity or for a non-unitary  \cite{DelbourgoTwisk}   massive Yang-Mills theory?

 Massive gravities, being bimetric, are also susceptible to causality problems if the relationship between the two metrics' null cones is not correct; sometimes it is not  \cite{Schmelzer}, 
at least not 
 without help \cite{MassiveGravity1}. Though matter sees only the effective curved metric and gravity only barely sees the flat background metric due to the smallness of the graviton mass, these theories are only Lorentz-covariant (or covariant under the 15-parameter conformal group in the case of massless spin $0$).  Thus the usual special relativistic arguments about superluminality in one frame implying backwards causation in another frame are applicable.

Whether or not massive variants of GTR prove to be viable theories, it is of considerable interest to  add gauge freedom artificially \cite{Schmelzer,Arkani,MassiveGravity1} to massive theories of gravity \cite{OP,FMS} in order
to compare Einstein's GTR, massive theories, and massive theories with gauge freedom.  This comparison, the analog of Maxwell \emph{vs.} Proca \emph{vs.} Stueckelberg comparison, juxtaposes the paradigmatic generally covariant theory (GTR), a theory formulation in which general covariance is broken totally but as simply as possible (massive GTR), and a formulation in which general covariance is in many respects restored artificially (gauged massive GTR).  If the massive theories prove theoretically healthy and empirically viable, then they also pose a case of underdetermination, as in electromagnetism.    Whether these theories are physically viable, however, is not crucial for present purposes. 

 The gauge freedom to be installed here is external, involving the crucial transport term $$\xi^{\alpha} \frac{\partial}{ \partial x^{\alpha}} g_{\mu\nu}$$ or the like from the Lie derivative \cite[p. 258]{Sundermeyer}, which has no analog for electromagnetic or Yang-Mills transformations, in which the derivative of the potential does not appear.  This sort of term  is responsible for issues of space-time point individuation in the hole argument. There have been various studies that introduce gauge compensation or so-called Stueckelberg fields for \emph{linear} theories of spin 2 (and perhaps also spin 0) fields 
\cite{DelbourgoSalam,HamamotoVector,HamamotoGauge1,HamamotoGauge2,DuffLS,Zinoviev}.   
Current work on higher-spin (that is, higher than the spin $2$ of gravity) fields often takes place in a ``gauge invariant'' formalism making use of gauge compensation fields \cite{BuchbinderKrykhtinFlat,BuchbinderKL}.  However, it is not at all clear what form the generalized Stueckelberg trick should take for nonlinear theories involving spin $2.$  Indeed one might worry that it would be horribly complicated and involve derivatives to all orders (thus being nonlocal), given the exponential form of the finite Yang-Mills gauge transformations and given the presence of derivatives of all orders in finite gauge transformations in gravitation \cite{Grishchuk}.  Fortunately the answer is available, painlessly, in a simple form \emph{via} parametrization \cite{Schmelzer,Arkani,MassiveGravity1}.



Making use of parametrization and some short-cuts, this section will show in effect what obtains from the application of a BFT-type procedure (modified as above for Lagrangian-friendliness) to a nonlinear massive theory of gravity, the FMS theory \cite{FMS,LogunovBook}. 
This theory is in some respects privileged over  other massive gravities \cite{OP,OPMassive2}, though  not by virtue of the universal coupling derivation, which applies to all of the Ogievetsky-Polubarinov theories \cite{MassiveGravity1} (\emph{pace} \cite{DeserMass}).  Making contact with parametrization, along with some known results involving the Poisson bracket algebra for GTR \cite{Sundermeyer}, obviates difficult calculations.   
 For present purposes the fact that the spin 2-spin 0 theories are related to Einstein's in the same way that Proca's is to Maxwell's trumps any concern (noted above) that spin 2-spin 0 theories might not be healthy theories of gravity. 
Converting a nonvariational object to a function of clock fields and their derivatives is an easy path to the sort of artificial general covariance   \cite{SchmelzerMass,Arkani,MassiveGravity1} that presumably would result from a more involved BFT-type procedure applied to massive GTR. The reduction back to the original formulation is accomplished readily by setting, for example, $X^M = x^{\mu}$ everywhere.  Thus nonvariational fields and clock fields appear to be inter-convertible.  However, if one has introduced gauge freedom in order to achieve some other goal, such as a consistent notion of causality \cite{MassiveGravity1}, then the condition $X^M = x^{\mu}$  might not be an allowed gauge condition.  Otherwise inter-convertibility seems to hold in general.

An additional philosophical reason to consider versions of gravity with a non-zero `graviton rest mass' is the way that they exemplify the importance, as discussed recently by Harvey Brown \cite{BrownPhysicalRelativity}, of giving physical explanations not (or not simply) in terms of space-time structure, but in terms of detailed physical laws.  Massive GTR \cite{OP,FMS} (forgetting its problems with positive energy and causality \cite{MassiveGravity1} for the moment) has the Poincar\'{e} group for its symmetry group, and yet matter `sees' only the effective curved metric $g_{\mu\nu}$, because the metric appearing in the matter action is the curved effective metric, not the flat background $\eta_{\mu\nu}$.  For sufficiently small graviton rest masses, in practice the flat background metric is very difficult to observe even in gravitational experiments in these theories, though it is certainly present in the field equations.  The fact that the symmetry group is the Poincar\'{e} group of STR might lead one to anticipate typical STR phenomenology, in which rods and clocks see the flat metric, but this anticipation is clearly seen to be false on examination of the matter action.  The ``space-time structure'' of these theories involves both a curved metric $g_{\mu\nu}$ and a flat metric $\eta_{\mu\nu}$, but this mere \emph{listing} of geometric objects leaves one with almost no idea what to expect from experiments. Why, as Brown might ask, does matter see the curved metric and not the flat metric?  Appeal to space-time structure here gives no explanation, but a glance at the matter action $S_{matter}[g_{\mu\nu}, u]$ (with matter denoted by $u$) gives a partial answer immediately: $\eta_{\mu\nu}$ is absent from the matter action and  appears only in the gravitational action.  (There remains the further question of why matter couples to $g_{\mu\nu}$ in such a fashion that material objects behave as rods and clocks for   $g_{\mu\nu}$, rather than relating to $g_{\mu\nu}$ in some more complicated fashion \cite{BrownPhysicalRelativity}.\footnote{I thank Katherine Brading for discussing this matter.})  These same points apply to massive \emph{scalar} gravity \cite{FreundNambu,PittsScalar}, which generalize  Nordstr\"{o}m's theory \cite{DeserHalpern},  with flat metric $\eta_{\mu\nu}$ and the merely conformally flat metric $g_{\mu\nu}$ conformally related to it. Matter sees only the conformally flat $g_{\mu\nu}$; gravity alone involves the flat metric  $\eta_{\mu\nu}$, because $\sqrt{-\eta}$ appears in the mass term for gravity, but nowhere else. Massive scalar gravity turns out to be wrong empirically, predicting no bending of light, but that does not matter for present purposes; unlike massive GTR, massive scalar gravity is free of worries about positive energy or causality.    The fact that explanations in terms of space-time structure appear to work so well for STR and GTR is due to the scarcity of geometric objects arguably pertaining to geometry (as opposed to matter) in the theory, yielding (almost) unique results for matter coupling and the behavior of the metric.    But given the possibility of writing theories such as massive gravities, or even theories in which different kinds of matter `see' radically different space-time structures, explaining physical phenomena in terms of mere space-time structure, without detailed recourse to the Lagrangian density, is typically an unsuccessful strategy. 
Poincar\'{e} already considered in principle the idea of a theory with two conformally related metrics as an argument for the conventionality of geometry \cite[pp. 88, 89]{PoincareFoundations}  \cite{BenMenahemPoincare}.  In the absence of any serious example, the idea fell by the wayside.  However, one need only have done to Nordstr\"{o}m's theory what Neumann and Seeliger had done to Newton's in the 1890s (see \cite{Pauli,North,NortonWoes}) in order to invent massive scalar gravity.  It would be interesting to rationally reconstruct the history of 20th century space-time philosophy with such examples imagined to have been available by the time that Eddington's enthusiastic endorsement of geometric empiricism  \cite{EddingtonSTG} appeared.


The treatment of the  FMS theory \cite{FMS} in Hamiltonian form will now be recalled \cite{DeserMass,MassiveGravity1}.  The  Lagrangian density is (apart from divergences and other unimportant terms and an overall factor) 
\begin{equation}
\mathcal{L}=\sqrt{-g}R(g)-m^2 \left(-\sqrt{-g}-\sqrt{-\eta}+
\frac{1}{2}\sqrt{-g}\,g^{\mu\nu}\eta_{\mu\nu}\right).
\end{equation}
This Lagrangian density includes a \emph{formal} cosmological constant term $\sqrt{-g}$ and an unimportant constant term $\sqrt{-\eta}$, but  the  term $-\frac{m^2}{2}\sqrt{-g}\,g^{\mu\nu}\eta_{\mu\nu}$ breaks the gauge symmetry by introducing preferred coordinates implicitly. This term cancels the linear term in the gravitational potential (construed as something like the difference between the curved and flat metrics), which is responsible for the peculiar cosmological constant behavior with a potential growing with distance \cite{FMS}.  The distinctive mass term phenomenology, with Yukawa exponential decay of the potentials, then arises from the \emph{quadratic part} of $\sqrt{-g}$ in the \emph{absence of the linear part}. It has been suggested from time to time, going back to Einstein \cite{EinsteinCosmological}, that the cosmological constant term itself introduces a mass term for gravity.  While that claim is false \cite{DeWittDToGaF,FMS,Treder,CooperstockTerm}, it is evident that the cosmological constant can indeed contribute to a mass term, if the linear term is cancelled.

Making the usual ADM  (3$+$1)-dimensional split \cite{MTW} of the curved metric $g_{\mu\nu}$, one uses for dynamical variables
the lapse function $N$ relating the  proper time to coordinate time, the  shift vector $\beta^i$ expressing how the spatial coordinate system moves over time, and a curved spatial metric $h_{ij}$
with the inverse $h^{ij}$ and determinant $h$. Letting $g^{\mu\nu}$ be the
inverse curved metric as usual, one has $g^{00}=-N^{-2}$, $g_{ij}=h_{ij}$, and $g_{0i}=h_{ij}\beta^j$. 
For temporary convenience I partly fix the coordinates to have $\eta_{00}=-1$ and $\eta_{0i}=0.$
The above Lagrangian density, after dropping a divergence, becomes
\begin{equation}
\mathcal{L}=N\sqrt{h}\left[R+K_{ab}K^{ab}-K^2+
m^2\left(1-\frac{h^{ij}\eta_{ij}}{2}\right)\right]+
m^2\left[\sqrt{-\eta}+\frac{\sqrt{h}}{2N}(\eta_{ij}\beta^i\beta^j-1)\right].
\end{equation}
The canonical momenta, as in GTR, are 
\begin{equation}
\pi^{ij}=\frac{\partial \mathcal{L}}{\partial h_{ij,0}}=\sqrt{h}(K^{ij}-h^{ij}K),\qquad
P_i=\frac{\partial \mathcal{L}}{\partial\beta^i_{,0}}=0,\qquad
P=\frac{\partial \mathcal{L}}{\partial N_{,0}}=0.
\end{equation}
The four vanishing canonical momenta are  primary constraints in  constrained dynamics \cite{Sundermeyer}.

Performing the generalized Legendre transformation and using the primary
constraints gives the canonical Hamiltonian density
\begin{equation}
\mathcal{H}=N\left[\mathcal{H}_0+m^2\sqrt{h}\left(\frac12 h^{ij}\eta_{ij}-1\right)\right]+
\beta^i \mathcal{H}_i-m^2\sqrt{-\eta}+
\frac{m^2\sqrt{h}}{2N}(1-\eta_{ij}\beta^i\beta^j),
\label{46}
\end{equation}
where, as usual,
$$
\mathcal{H}_0=\frac1{\sqrt{h}}\left(\pi^{ij}\pi_{ij}-
\frac12\pi^2\right)-\sqrt{h}\,R,\qquad\mathcal{H}_i=-2D_j\pi^j_i,
$$
and $D_j$ is the three-dimensional torsion-free covariant derivative
compatible with $h_{ij}$. Setting  $m=0$ recovers the usual GTR form that is purely
a sum of constraints, but $m\neq 0$ destroys that form and leads to six, not
two, degrees of freedom at each point in space. The zeroth-order term
$-m^2\sqrt{-\eta}$ has been retained to give Minkowski space-time zero energy. 
The secondary constraints are obtained in effect by varying the lapse $N$
and shift vector $\beta^i$, yielding  the
modified Hamiltonian constraint
\begin{equation}
\frac{\partial \mathcal{H}}{\partial N}=\mathcal{H}_0+m^2\sqrt{h}
\left(-1+\frac12h^{ij}\eta_{ij}\right)-
\frac{m^2\sqrt{h}}{2N^2}(1-\eta_{ij}\beta^i\beta^j)=0
\end{equation}
and the modified momentum constraint
\begin{equation}
\frac{\partial\mathcal{H}}{\partial\beta^i}=\mathcal{H}_i-\frac{m^2\sqrt{h}}{N}\eta_{ij}\beta^j=0.
\end{equation}
These constraints are second-class \cite{PittsQG05}. Thus this theory is  a gravitational analog of Proca's massive electromagnetism.

With the BFT procedure as amended above   regarding the primary constraints  and the boundary conditions on the gauge compensation fields, one should be able to  install artificial gauge freedom in the FMS theory in much the same fashion as in Proca's electromagnetism, apart from computational difficulty.  The task is easy, however, if one makes use of parametrization.  One  need only make the transformation  $\eta_{\mu\nu} \rightarrow X^{A},_{\mu} \eta_{AB} X^{B},_{\nu}$ in a single piece of the mass term, $-\frac{m^2}{2}\sqrt{-g}\,g^{\mu\nu}\eta_{\mu\nu}$, the piece that breaks the gauge symmetry,  and the work is nearly finished. 
Apart from the constant term $\sqrt{-\eta}$ which one leaves  alone, one obtains a Lagrangian formally the same as GTR with four minimally coupled massless scalar fields and a cosmological constant. This Lagrangian density is manifestly coordinate-invariant and has no nonvariational fields in the action principle, so GTR-type generalized Bianchi identities relating the Euler-Lagrange equations follow.  When the clock fields are expanded in a background piece and perturbing gauge compensation field, the result at lowest order  resembles the Stueckelberg trick. The fact that the clock fields look formally like matter fields in the Lagrangian density is a reminder of the flexibility in deciding which fields pertain to gravity and space-time and which do not.

While the success of the installation of gauge freedom is already now evident at the Lagrangian level, it can be displayed perspicuously using the Hamiltonian also.  
 Applying the Dirac-Bergmann constrained dynamics procedure to the result, one readily sees that all constraints are now first-class.  The cosmological constant term does not affect the Poisson bracket algebra of constraints; neither do the minimally coupled scalar fields \cite[p. 253]{Sundermeyer}.    That one of the scalar fields has negative energy is relevant to the viability of the theory, but not to the Poisson bracket algebra. Contrary to the usual BFT procedure,  I require the primary constraints $P$ and $P_i$ and the other old momenta to suffer no change while gauge freedom is installed. The new gauge compensation fields, which are clock fields $X^{A},$\footnote{In this case the boundary condition of trivializing the new fields means not setting them to $0,$ but setting them to equal the coordinates $x^{\alpha}.$}
 have canonical momenta $\pi_A = -m^2 \sqrt{-g} g^{0\mu} X^{B},_{\mu} \eta_{BA}.$
 Gauge freedom is installed by adding a term to the Hamiltonian density, 
\begin{equation} \Delta \mathcal{H} = N \mathcal{H}_{0s} + \beta^{i} \mathcal{H}_{is} -\frac{m^2}{2}\sqrt{-g}\,g^{\mu\nu}\eta_{\mu\nu},
\end{equation}
where $$\mathcal{H}_{0s} = \frac{\pi_A \eta^{AB} \pi_B}{2 m^2 \sqrt{h} } + \frac{m^2}{2} X^{A},_i \eta_{AB} X^B,_j h^{ij} \sqrt{h}$$ and $$ \mathcal{H}_{is} = X^A,_i \pi_A.$$
The altered Hamiltonian with the unchanged primary constraints gives the new secondary constraints.  They have the same Poisson bracket algebra as vacuum GTR, one sees without calculation, because neither the cosmological constant, nor the minimally coupled scalar fields, nor the constant term $\sqrt{-\eta}$ changes the algebra.

Studying the FMS massive theory of gravity and its parametrized variant has illuminated questions involving empirical equivalence in much the way that studying the Proca and Stueckelberg massive electromagnetisms  illuminated the question of empirical equivalence for electromagnetism.  This parallel reflects and illustrates and deep technical and conceptual similarities between Maxwell's electromagnetism and Yang-Mills theories on the one hand, and Einstein's GTR on the other.  The FMS theory is distinguished among the 2-parameter family of massive gravities \cite{OP,OPMassive2} by its containing the flat metric $\eta_{\mu\nu}$ to the first power only, not higher powers or the inverse metric $\eta^{\mu\nu}$ or the determinant $\sqrt{-\eta}$ in the crucial symmetry-breaking term.  It is this fact that, after parametrization, yields the Lagrangian that is formally just four massless scalar fields (one with the ``wrong'' sign) with the standard kinetic term.  Parametrization of any of the other OP 
massive gravities will result in higher powers of $\partial X$.  Whether the resulting exotic kinetic terms fall within the scope of minimally coupled matter (as understood at the time of (\cite[p. 253]{Sundermeyer}) is unclear.  Moreover, given the many ways that the lapse $N$ and shift $\beta^i$ and the field derivatives $\partial X$ enter the Lagrangian densities for most OP theories, finding the Hamiltonian could be challenging.  Thus ascertaining by direct calculation whether the Poisson bracket algebra has changed could be difficult.  
Fortunately it is clear on Lagrangian grounds that the replacement $\eta_{\mu\nu} \rightarrow X^A,_{\mu} \eta_{AB} X^B,_{\nu}$  has the desired effect in these theories also.  The manifest coordinate covariance of the Lagrangian density, a weight $1$ scalar density with no nonvariational fields such as $\eta_{\mu\nu}$ in  the Euler-Lagrange equations, indicates that there will be generalized Bianchi identities \cite{Sundermeyer} much as in GTR.  One also knows how to count gauge invariances and degrees of freedom directly from the Lagrangian \cite{HenneauxTeitelboimZanelli,ShepleyEvolutionary}; doing so gives the expected results. 


Studying the FMS massive theory of gravity and its parametrized variant has shed light on general covariance.
Massive variants of GTR are interesting because their Lagrangian densities are, except for one term key term, those of GTR.  Thus one can isolate most or all the phenomena of substantive general covariance or its violation; the absence or presence of a mass term implies the presence or absence of substantive general covariance, respectively (assuming that GTR exemplifies it, an assumption that is not quite so obvious after the Geroch-Giulini $\sqrt{-g}$ counterexample to the Anderson-Friedman absolute objects program).  Installing artificial gauge freedom by parametrization at least formally restores many features of GTR that one might associate with substantive general covariance:  gauge freedom, first-class constraints, a Hamiltonian that is a sum of constraints (perhaps apart from a constant in this case), point individuation questions such as appear in the hole argument, the absence of non-variational fields in the Lagrangian density (apart from a constant term), \emph{etc.}   However, parametrized GTR's gauge freedom is clearly artificial.  Apart   from causality worries that appear to problematize gauge-fixing \cite{MassiveGravity1} to recover the original FMS formulation, it would be exceedingly natural to regard the parametrized and non-gauge formulations as the same theory, much as one might identify the Proca and Stueckelberg formulations of massive electromagnetism.  That parametrization of a non-gauge theory such as FMS massive gravity can mimic so many features of GTR that one associates with substantive general covariance tends to confirm that the presence or absence of clock fields is perhaps the best criterion for deciding whether a formally generally covariant theory is \emph{merely} formally generally covariant, or is substantively so.

The use made of artificial gauge freedom in relation to massive GTR is, I believe, novel.  The original formulation has a subtle but important physical defect pertaining to causality.  By installing \emph{formal} artificial gauge freedom (neglecting for the moment whether the resulting configurations are all equivalent), one finds that some but not all  configurations in the enlarged formulation are free of the defect.  Taking the core idea of gauge freedom to involve physical equivalence, and letting the group structure drop by the wayside as a consequence, one stipulates that gauge transformations are those transformations that relate physically equivalent configurations, which in these case means configurations with the appropriate causality properties.  One therefore obtains a gauge groupoid (in the sense of Brandt), not a gauge group: the allowed transformations depend on the configuration in question, so not every gauge transformation can be composed with every other.  While in principle one ought to be able to gauge-fix the resulting formulation, simply turning off the new fields by setting them to trivial values is not permitted.  Sometimes apparently surplus structure isn't so surplus after all, even though it remains unobservable.


\section{Generalized BFT Conversion,  Unconstrained Theories, Parametrization and a Further Generalized Kretschmann Objection}

The Einstein-Kretschmann debate over the significance of general covariance  inspires a generalized Kretschmann objection to the idea that any sort of gauge freedom is physically significant \cite{CMartinSymmetries}.  While philosophers of physics probably generally believe that one can add gauge symmetries to theory formulations that lack them, there has been rather little discussion of how to do so explicitly in an interesting way.  In BFT conversion, we now have in hand an algorithm for mounting a generalized Kretschmann objection.

For philosophical and physical purposes, there might be considerable interest in various generalizations of the BFT-type procedure. 
Why must one be  interested only in  converting a theory formulation with only second-class constraints into a theory with that same number of first-class constraints and no second-class constraints?  There are physically interesting cases of theories naturally formulated 
 with first-  as well as second-class constraints \cite{ParkPark,Monemzadeh,KimParkYoonEW}; clearly a generalization of the BFT conversion algorithm is required to accommodate such cases.  This generalization, however, is hardly ground-breaking conceptually.  There might also be reasons, however, to perform only a partial conversion, leaving some second-class constraints and some first-class constraints.  Such a procedure might be worthwhile in cases, perhaps such as massive gravity, where installation of gauge-freedom is made to satisfy an important physical principle such as null-cone causality, but one wants to stay as close to the true degrees of freedom as possible.  If  some 1960s writers on massive gravity \cite{OP,FMS} had noticed that their theories were acausal \cite{Schmelzer,MassiveGravity1}, perhaps they, writing before turning second-class constraints into first-class constraints came to seem like a good idea in general, would have sought such a procedure.  
There are also reasons for studying unconstrained theories parametrized with clock fields,  such as a massive scalar field satisfying the Klein-Gordon equation in Minkowski space-time \cite{Kuchar73}.  The BFT conversion algorithm takes for granted that there are second-class constraints that one wants to  turn into first-class constraints, as with massive gravity \cite{Schmelzer,Arkani,MassiveGravity1}; the BFT procedure, strictly construed, is simply inapplicable to unconstrained theories.  One can imagine even more general possibilities, such as leaving some second-class constraints untouched while adding gauge freedom elsewhere in a theory, such as by parametrizing the Proca theory.  Kucha\v{r} and C. L. Stone have taken Maxwell's electromagnetism, a theory with first-class constraints, and parametrized it \cite{KucharStoneParam}.  The general idea of adding artificial gauge freedom in fact is not tied essentially to the presence of second-class constraints, or of any constraints at all, in the initial theory formulation.

A sufficiently  general sort of procedure that adds gauge freedom seems to be the following.  Take a theory formulated in terms of a Hamiltonian (perhaps starting from a Lagrangian); there need not be any constraints of any kind. Add a new term (chosen by an educated guess or with sufficient generality) to the Hamiltonian and perhaps some new constraints (perhaps obtained by starting with the Lagrangian).  Run the Dirac-Bergmann stabilization algorithm to find all the constraints.  Gauge freedom has indeed been added if there are  at least some first-class constraint(s) that have no first-class ancestors in the original formulation (though there might be second-class ancestors), while retaining (perhaps under modification) whatever first-class constraints were present in the initial formulation as well.  (The second clause is to ensure that one does not lose any gauge symmetries in the process, because that seems to violate the spirit of adding gauge freedom.) Confirm  that the new formulation has the same number of degrees of freedom as the original.  Verify that the new formulation can be gauge-fixed into the original one when the new fields reduce to some trivial configuration.  The successful implementation of this sort of procedure installs gauge freedom even in theories that perhaps had no constraints initially, such as if one parametrizes the theory of a massive or massless scalar field. Parametrizing Proca's electromagnetism, which starts with second-class constraints in the electromagnetic sector and then acquires first-class constraints pertaining to space-time, should also fit within this framework. In this fashion one can subsume both the parametrization process and the BFT conversion algorithm into a more general procedure for adding gauge freedom.
Parametrization adds external gauge symmetry to theories that might or might not have any constraints.  The BFT procedure adds gauge freedom, whether internal or external, to theories with second-class constraints.  The more general procedure just outlined adds gauge freedom,  whether internal or external or both, to theories that might or might not have any constraints.  Both the BFT procedure and parametrization are applicable in the case of massive gravity \cite{OP,FMS}, which can be viewed as having a geometrical symmetry broken by the mass term.  The demonstration  above 
that parametrization of FMS massive gravity  yielded just the sort of result that one expects from BFT (modified concerning the  boundary conditions and the  distinction between coordinates and momenta)   confirms the utility of subsuming both the BFT and parametrization procedures into a more general field-theoretic technique.


\section{Artificial Gauge Freedom and Philosophy}

\subsection{Post-Positivist  Views on Underdetermination: Glymour and Quine}

	Relatively recent (post-positivist)  discussions of underdetermination and empirical equivalence are not entirely unaware of the possibility of adding descriptive redundancy to theories.  According to  Sklar, \begin{quote} [t]ypically one can generate alternative theories saving the phenomena by some process which introduces into a theory otiose elements whose place in the theory ``cancels out.''  Most interesting, of course, are the historical cases where the theory with the otiose elements came first and where it was an important scientific discovery that one could eliminate them by a conceptual revision. \cite[p. 62]{SklarNoumena}  \end{quote}
As Sklar's comment suggests, there is no expectation  that introducing such ``otiose'' elements might be scientifically productive. Van Fraassen's versions of Newtonian mechanics with different velocities for the center of mass of the solar system (re)introduce a quantity that is unobservable due to the relativity of motion in the theory \cite{vanFraassenSI,LaudanLeplin}.  While one might regard the Stueckelberg,  BFT and parametrization technologies as introducing a generalized relativity of motion---that is, of the evolution of fields---these philosophical discussions do not recognize the  technical sophistication that such introductions might involve, and indeed already had involved, in the physics literature.  
Recent discussions of empirical equivalence  often imply one or more  claims which now seem doubtful in certain important physical cases.  First, the formulation with the extra entities is often taken to be a different theory from the original leaner formulation.  Second, the formulation with the extra entities is thought to have no practical advantage over the original leaner formulation. Third, the formulation with the extra entities is thought to have no conceptual advantage over the original leaner formulation. After commenting  on the presence of some or all of these themes in works by Glymour and  Quine,  I will recall how each of these claims is at variance with the  beliefs and practices of contemporary physicists working on certain problems.

Quine mentions briefly a possibility that might accommodate the transition from  Proca theories to  Stueckelberg theories:
\begin{quote} \ldots  suppose we had an adequate theory of nature, and then we were to add to it some gratuitous further sentences that had no effect on its empirical content.  By ringing changes on these excrescences we might get alternative theories, logically incompatible, yet always empirically equivalent.  This gratuitous branching of theories would be of no interest to the thesis of under-determination, since the adequate original theory was itself logically compatible with each one of these gratuitous extensions; they were incompatible only with one another. \cite[p. 323]{QuineEquivalent} \end{quote}
Taking some Proca theory for the moment as the ``adequate theory of nature''---obviously not a realistic claim, but it does even quantize well---can one take the Stueckelberg formulation merely to add ``some gratuitous further sentences''?  If one makes the field redefinition 
$$ \tilde{A}_{\mu} =_{def} A_\mu + \partial_\mu \phi,$$
one can rewrite the Stueckelberg Lagrangian density 
$$ 
\mathcal{L}_{s} = -\frac{1}{4} F_{\mu\nu}F^{\mu\nu} - \frac{m^2}{2} (A_\mu + \partial_\mu \phi) (A^\mu+ \partial^\mu \phi).$$ 
in terms of $\tilde{A}_{\mu}$ and $\phi,$ but $\phi$ no longer appears in the Lagrangian. 
While field redefinitions with derivatives are not to be trusted without question (\emph{c.f.} the proof for algebraic redefinitions \cite{Schouten}), this one seems harmless.  
 The Euler-Lagrange equations for $\tilde{A}_{\mu}$ make no reference to $\phi$ (because the expression $A_\mu + \partial_\mu \phi$ in the mass term yields the same field strength $F_{\mu\nu}$ as results from $A_\mu$) 
 and are identical to those of the Proca theory (apart from the typography of $\tilde{A}_{\mu}$).  
 This field redefinition can be taken as a  reconstrual of predicates that Quine allows in testing for  theoretical equivalence.  The new field equation for $\phi$ is vacuous, so one can prescribe $\phi$ freely (fix the gauge): the extra field $\phi$ appears solely in ``some gratuitous further sentences'' with no empirical content.  Thus the application of Stueckelberg's trick to a Proca formulation fits within Quine's discussion.\footnote{An analogous  procedure appears to work for Yang-Mills theories \cite{Ruegg}, though that is not always obvious \cite{YMembed}. One should distinguish between an Abelian-like conversion $A^{a}_{\mu} \rightarrow A^{a}_{\mu} + \partial_{\mu} \phi^a$ of the Yang-Mills mass term used by some authors \cite{UmezawaKamefuchi,Salam} on the one hand, and  the conversion $A^{a}_{\mu} \rightarrow A^{a}_{\mu} + D_{\mu} \phi^a$ resembling a non-Abelian gauge transformation on the other \cite{GrosseKnetter}. 
 The latter has been criticized for not actually converting all second-class constraints into first-class constraints \cite{BanerjeeGhosh}, contrary to claims made about it. Given the consequences for counting degrees of freedom and given the interrelation between Hamiltonian and Lagrangian symmetries \cite{HenneauxTeitelboimZanelli}, this controversy is surprising, but I have not attempted to resolve it.    
For parametrized massive GTR, the relevant field redefinition is
$g^{AB}=_{def} g^{\mu\nu} X^{A},_{\mu} X^{B},_{\nu}$ and the like; using the preferred coordinates $X^{A}$ to define  the action integral makes the arbitrary label coordinates $x^{\mu}$ disappear from the action.}  While for Quine the Stueckelberg formulation is logically compatible with the Proca formulation, the Stueckelberg formulation (even after the field redefinition that reconstrues the predicates) is a different theory from Proca's:  the Stueckelberg theory posits the existence and certain behavior of the field $\phi$, whereas the Proca theory does not.   The connotations of ``gratuitous further sentences'' and ``ringing changes on these excrescences'' indicate that Quine would have anticipated nothing practically useful or conceptually insightful to result from the Stueckelberg trick.\footnote{Quine's relativism about ontology and general pro-science naturalistic form of philosophy \cite{QuineOntological} are more helpful than the remarks quoted above, as viewed from particle physics.  Confronted with the discussion to follow, he might conclude that at the quantum level,   it is Stueckelberg rather than Proca that gives an adequate theory, in which case my criticisms become largely inapplicable.  I thank Don Howard for discussion.}

Clark Glymour's discussion of the underdetermination of geometry (or otherwise) can be substantially carried over to the case of artificial gauge freedom.  Glymour, responding to Reichenbach, invites the reader to 
\begin{quote}
[s]uppose you find yourself teaching high school physics, Newtonian mechanics in fact.  Suppose further than a bright and articulate student named Hans one day announces that he has an alternative theory which is absolutely as good as Newtonian theory, and there is no reason to prefer Newton's theory of his.  According to his theory, there are two distinct quantities, gorce and morce; the sum of gorce and morce acts exactly as Newtonian force does.  Thus the sum of gorce and morce acting on a body is equal to the mass of the body times its acceleration, and so on.  Hans demands to know why there is not quite as much reason to believe his theory as to believe Newton's.  What do you answer?

I should tell him something like this.  His theory is merely an extension of Newton's.  If he admits that an algebraic combination of quantities is a quantity, then his theory is committed to the existence of a quantity, the sum of gorce and morce, which has all of the features of Newtonian force, and for which there is exactly the evidence there is for Newtonian forces.  But in addition his theory claims this this quantity is the sum of two distinct quantities, gorce and morce.  However, there is no evidence at all for this additional hypothesis, and Newton's theory is therefore to be preferred.\ldots

The gorce plus morce theory is obtained by replacing ``force'' wherever it occurs in Newtonian hypotheses by ``gorce plus morce'', and by further claiming that gorce and morce are distinct quantities neither of which is always zero.\ldots  [T]he computations which give values for force will not give values for gorce or for morce, but only for the sum of gorce and morce.  Indeed, in general if we have a set of simultaneous equations such that using these equations, values for some of the variables in the equations may be determined from values of other variables, then values for the new variables will not be determined.\ldots Implicit in the discussion is a certain articulation of the principle that we prefer a theory with fewer untested hypotheses to one with more untested hypotheses.  \cite{GlymourEpist}. \end{quote}
Whether this analogy accurately captures the main issues involved in the question of the conventionality of geometry may be doubted \cite{NullCones1}, but that need not concern us now. It fits  well enough  the transition from a Proca formulation to a Stueckelberg formulation of massive electromagnetism (apart from the assumption of a merely  algebraic   relation among the fields), and \emph{mutatis mutandis} the general process of installing artificial gauge freedom.  It is evident that Glymour considers the gorce plus morce theory distinct from Newton's theory, practically in no way advantageous to Newton's theory, and conceptually inferior to Newton's theory.  By parity of reasoning, one might expect a similar verdict from him on the Stueckelberg formulation of massive electromagnetism and on the results of BFT conversion or parametrization; at the least, one can imagine that many philosophers, confronted with this example, might reason analogously to what Glymour says about gorce plus morce.

Quine and Glymour (and they seem not to be unusual among philosophers) take the  formulations with additional unobservable entities not only to be different (though compatible) theories from the leaner originals, but also to be practically and conceptually inferior to the leaner originals.  By contrast, it is overwhelmingly taken for granted in the  physics literature that the Stueckelberg formulation is a formulation of the very same theory as the Proca formulation.  More generally, it is overwhelmingly taken for granted  that formulations obtained using the BFT algorithm or parametrization are formulations of the same theory after the surgery as before.  With the possible exception of A. A. Logunov and his school, no physicist (to my knowledge) regards gauge-fixing as producing a numerically distinct theory.   When gauge freedom has been installed artificially, as in the Stueckelberg, BFT and parametrization cases, one can eliminate the extra fields, the ``otiose elements'' or ``excrescences,'' by gauge-fixing.  Whereas physicists agree that the result is the very same theory as before gauge-fixing, notable  philosophers of science have been committed to the view that a different and superior theory is  produced by gauge-fixing.  Whereas such philosophers of science would expect the formulations with artificial gauge freedom to be practically inferior to their leaner ancestors, physicists find artificial gauge freedom useful in taking the limit of vanishing photon or other particle mass \cite{Zinoviev},  studying higher spin fields \cite{BuchbinderKrykhtinFlat},  avoiding the technical challenges sometimes present with using the true degrees of freedom by installing gauge freedom artificially   \cite{HenneauxTeitelboim,BanerjeeGhosh,ParkPark}, and proving desirable properties of a physical theory in different gauges, such as unitarity in one gauge and renormalizability in another \cite[chapter 21]{WeinbergQFT2}\cite[chapter 10]{Kaku}.  On the other hand, if one individuated theories as strictly as do Quine and Glymour, one would have (at least) a theory that had  gauge freedom but was not obviously unitary or renormalizable, an empirically equivalent but distinct theory that had no gauge freedom and that was unitary but not obviously renormalizable, and a third empirically equivalent but distinct theory that lacked gauge freedom and that was renormalizable but not obviously unitary.
 If one can regard the tetrad-spinor formalism as obtained from the OP spinor formalism by the installation of artificial gauge freedom \cite[p. 234]{GatesGrisaruRocekSiegel} (perhaps in a Lakatosian rational reconstruction of the history of spinors in curved space-time),
then the linearity of the tetrad-spinor formalism is another  practical benefit from artificial gauge freedom.
Parametrizing special relativistic theories has been fruitful conceptually in revealing similarities between GTR and parametrized theories, as the discussion of the suggestion that GTR is ``already parametrized'' demonstrated.  Parametrizing massive GTR also has the advantage---it is unclear to what degree it is practical \emph{vs.} conceptual---of taking acausal theories and giving them some hope of causality \cite{MassiveGravity1}. While this last application does not strictly contradict Quine or Glymour, who assumed that one had an adequate theory at the beginning, it does exemplify the importance of artificial gauge freedom, contrary to what they presumably would have expected. 

 How is it that  philosophers of science in the 1970s-80s tended to disagree  with physicists on the value of artificial gauge freedom?  One explanation is that the relevant physics has become prominent only rather recently, perhaps too recently to have influenced the philosophical works discussed; newer works will be considered shortly.  A deeper reason, however, is that physicists, often tacitly, ascribe to physical theories an  ontology that differs from the \emph{prima facie} ontology indicated by the fields in the Lagrangian density.  To be specific, physicists take for granted, and sometimes assert, that the real is the invariant,\footnote{Don Howard has  called attention to this  theme \cite{HowardIdeal}.} where the relevant symmetry group (or groupoid \cite{Hahn,Renault}) is picked out (primarily) by the symmetries of the action.  That only gauge- and coordinate-invariant entities are considered real, and thus part of a theory's ontology, is a limited application of the principle of the identity of indiscernibles.  Though that principle is rarely invoked by name among physicists, the idea is ubiquitous.  Thus the real ontology of the Stueckelberg formulation does not include the extra scalar field $\phi$, because $\phi$ is ``mere gauge'' (indeed gauge-equivalent to $0$) and hence unreal or ``unphysical.''   While the identity of indiscernibles is far from obscure among philosophers,  recent philosophers of science who are inclined toward scientific realism have avoided invoking  the principle in some contexts where physicists almost unanimously invoke it.  One might take the modern reemergence of the hole argument in GTR \cite{HoleStory} as evidence of this  realism among philosophers of science, though doubtless modern mathematicians' fondness for active diffeomorphisms (which seem to require mathematical haecceities) has also contributed.   To a large degree, modern physical theories dictate their own ontologies, indicating clearly by the presence of gauge freedom---especially artificial gauge freedom, which typically can be removed without spoiling manifest locality \cite[p. 190]{BelotPoP}---when the naive ontology is not correct.  The venerable  tradition  of trying to express scientific theories in terms of logic and set theory, as is evident in Quine \cite{QuineEquivalent} among others, is rather remote from the language of contemporary physics \cite{LadymanRoss}. Thus Quine regards as equivalent two formulations that are logically equivalent after reconstrual of predicates, but a formulation that adds superfluous entities with arbitrary values produces, in his view, a different and inferior theory.    While explicitly eliminating superfluous fields is important for some tasks in the philosophy of physics, such as the Anderson-Friedman absolute objects program discussed elsewhere \cite{FriedmanJones,PittsPhilDiss}, to some degree Lagrangian field theory takes care of itself ontologically with the help of the principle that the real is the invariant.  I observe that a de-emphasis on pinning down the unobservable things posited by scientific theories and greater emphasis on mathematics is broadly in harmony with structural realism \cite{Worrall}.


\subsection{Artificial Gauge Freedom and More Recent  Views on Surplus Structure and Gauge Freedom}

In the last few decades, as the philosophy of physics has flourished, philosophers have increasingly wrestled with the significance of gauge freedom and what Michael Redhead has called ``surplus structure'' \cite{RedheadSymmetryTheories,RedheadQFT,RedheadInterpretation}, a part of the formal structure without any claimed referent in the world.  Given that gauge transformations make distinctions that correspond to no observable difference, and plausibly to no real difference of any sort, one might be tempted to think that reduction to eliminate surplus structure in general, and gauge freedom in particular, would be the obvious move. 
 Such a move finds defenders, especially among those who sympathize with Leibniz's relationalism and his principle of the identity of indiscernibles. Gordon Belot has discussed these matters in detail \cite{BelotGeometry,BelotGauge,BelotPoP}.  As noted above, there are significant technical drawbacks to such reductions, such as obscuring spatial locality (as well as Lorentz invariance). Thus one might be inclined to keep the surplus structure and gauge symmetry in place.  
Less common among philosophers, however, is the idea that there might be good reasons to \emph{expand} the collection of fields used to describe the system.  Belot, indeed, has noted some examples from fluid mechanics where expanding the collection of fields is advantageous \cite{BelotGauge};  it is unclear, given the well-known awkwardness of variational formulations of fluid mechanics, for example, whether this suggestion should be expected to hold for more plausibly fundamental theories, however. 
Redhead \cite{RedheadInterpretation} and Dean Rickles \cite{RicklesReduceEnlarge} have noted that BRST symmetry, a way of enlarging the space of fields to include a rigid fermionic transformation that manages to encode a local gauge symmetry, can be useful. Thus the lesson that enlarging the collection of fields can be beneficial is not quite unknown.   BFT conversion and its proposed generalizations above, however, differ from BRST symmetry.  
The notion of adding local gauge symmetries, which we have seen to be common and fruitful in physics, merits philosophers' continued attention.


\section{Conclusion}

Whereas the Anderson-Friedman absolute objects program gives the ``wrong'' answer for the general covariance of GTR because of $\sqrt{-g}$ \cite{FriedmanJones,GiuliniAbsolute}, and testing for fields not varied in the action principle appears to give the right answer for perhaps a wrong reason \cite{PittsPhilDiss}, converting fields not varied in an action principle to functions of clock fields and their derivatives appears to give the expected  list of substantively generally covariant theories  for plausibly the right reason.  This proposal, which is largely inspired by the claim that GTR is already parametrized, bears considerable resemblance to the pre-Kretschmann views of Einstein on GTR's distinctively lacking preferred coordinates.  One might then view clock fields as a stratagem to deploy in a Kretschmann-type objection that  general covariance is trivial or purely formal, while taking the absence of clock fields in GTR to show that it, unlike STR, has no preferred coordinates and \emph{ipso facto} is substantively generally covariant.  Thus both trivial and substantive general covariance admit clear analyses in terms of clock fields. The study of massive GTR confirms the utility of using the absence or presence of clock fields as the criterion for substantive general covariance.

 The discussion of clock fields in the context of general covariance turns out to involve a special case of artificial gauge freedom, so  the same concept sheds light on general covariance in the philosophy of physics and on underdetermination in the general philosophy of science.  A Lagrangian-friendly modification of the BFT formalism for converting second-class constraints into first-class constraints was suggested above; then one immediately arrives at the Stueckelberg formulation starting from the Proca theories, and presumably directly arrives at parametrized massive GTR when starting with massive GTR.   The physics literature on artificial gauge freedom also proves to admit considerable generalization to cases with no second-class constraints and perhaps no constraints at all; one need not convert all second-class constraints into first-class constraints, either, depending on one's goals.  These techniques provide algorithms for making generalized Kretschmann-type objections, to the effect that gauge symmetries are trivial or merely formal. 

In addition to providing resources to formulate and generate generalized Kretschmann objections, these techniques might help to resolve them.  Induction over special features of massive electromagnetism, Yang-Mills theories, and GTR in the formulations with artificial gauge freedom suggests some criteria for distinguishing artificial gauge freedom from the `natural' gauge freedom in the massless cases (apart from the obvious presence of a mass term quadratic in the potential, which might not generalize).  For the massive variants with artificial gauge freedom, the gauge parameters appear algebraically, so the new gauge compensation fields have redundant field equations.  For the theories with natural gauge freedom, eliminating gauge freedom tends to produce nonlocal formulations, whereas the massive theories with artificial gauge freedom can be gauge-fixed (apart from causality worries for massive gravity) back into local field theories with no gauge freedom.  Whether these differences are the correct criteria or not, they suggest that there might exist relevant differences that permit distinguishing artificial gauge freedom from the more familiar natural sort, in much the way that the claim that GTR is already parametrized suggests
that substantive general covariance is formal general covariance achieved without clock fields. Descriptive redundancy need not be a nuisance.  In the present philosophical contexts, as well as in physics, it is a resource:   artificial gauge freedom sheds light not only on the question of empirically equivalent theories, but also on general covariance.  

\section{Acknowledgments} I thank Nick Huggett 
 for helpful comments.



\end{document}